\documentclass[12pt,letterpaper]{article}
\usepackage[a4paper, total={7in, 10in}]{geometry}

\usepackage{graphicx}
\usepackage{helvet}
\usepackage{authblk}
\usepackage{hyperref}
\usepackage{amsmath} 
\usepackage{amssymb} 
\usepackage{orcidlink} 
\usepackage[super,comma,sort&compress]  
   {natbib}

\makeatletter
\renewcommand{\maketitle}{\bgroup\setlength{\parindent}{0pt}
\begin{flushleft}
  \textbf{\@title}
  
  \@author
\end{flushleft}\egroup}
\makeatother
\newcommand{\red}[1]{\textcolor{red}{#1}}

\newcommand{\ncoll}{n_{\rm coll}}

\begin{document}
\title{Observation of Universal Expansion Anisotropy from Cold Atoms to Hot Quark-Gluon Plasma}
\date{\today}





\author[1,2,*,\orcidlink{0000-0003-1329-9139}]{Ke Li}
\author[1,2,\orcidlink{0000-0002-8250-176X}]{Hong-Fang Song}
\author[1,2,\orcidlink{0000-0002-6377-9424}]{Hao-Jie Xu}
\author[1,2,\orcidlink{0000-0003-2621-4376}]{Yu-Liang Sun}
\author[1,2,3,4,**,\orcidlink{0000-0002-8313-0809}]{Fuqiang Wang}


\affil[1]{School of Science, Huzhou University, Huzhou, Zhejiang 313000, China}
\affil[2]{Strong-Coupling Physics International Research Laboratory (SPiRL), Huzhou University, Huzhou, Zhejiang 313000, China}
\affil[3]{Department of Physics and Astronomy, Purdue University, West Lafayette, Indiana 47907, USA}
\affil[4]{Lead contact}


\affil[*]{Correspondence: like99@zjhu.edu.cn}
\affil[**]{Correspondence: fqwang@purdue.edu}
\maketitle
\section*{SUMMARY}
Azimuthal anisotropy has been ubiquitously observed in high-energy nuclear (heavy-ion) collisions, perceived to require hydrodynamic interactions. This work reports a study of anisotropic expansion of cold $^{6}$Li Fermi gases, released from anisotropic potential traps, under tunable interaction strength by a magnetic field. A universal scaling of the momentum anisotropy response to geometry is observed for the first time between cold-atom and heavy-ion systems as a function of opacity--the average number of collisions per particle ($\ncoll$), despite their vast differences in scale and physics. The anisotropy response increases quickly at small opacity, without the need of a large amount of interactions, and shows no sign of saturation in the observed range, with an approximate power-law dependence of $\sqrt{\ncoll}$, characteristic of random walks. This universality potentially unifies a variety of vastly different physical systems, from dilute atomic gases to the hot and dense quark-gluon plasma of the early universe.

\section*{KEYWORDS}
heavy-ion collisions, cold atoms, azimuthal anisotropy, universal expansion

\section*{INTRODUCTION}
It is believed that the universe started with a big bang singularity—a vacuum where all charge quantum numbers are zero. The universe expanded, cooled down, and turned from a soup of quarks and gluons into particles like protons and neutrons at a temperature around $10^{12}$ Kelvin (or $\sim 200$~MeV) approximately $10~\mu$s after the big bang~\cite{Collins:1974ky,Boyanovsky:2006bf}. 
Protons and neutrons make atomic nuclei and all visible matter of stars and galaxies we see today. To study the state of the early universe, physicists collide heavy nuclei at speeds over $99.99\%$ of the speed of light, at the Relativistic Heavy-Ion Collider (RHIC) at Brookhaven National Laboratory, New York and the Large Hadron Collider (LHC) at CERN, Geneva, to create the state similar to that in the early universe, called the quark-gluon plasma (QGP)~\cite{Shuryak:1978ij,Muller:2006ee,Busza:2018rrf,Muller:2012zq,Roland:2014jsa}. 
Much like the early universe, the QGP expands, cools down, and undergoes a phase transition to a system of hadrons. Contrary to the early expectation of a free or weakly interacting gas of quarks and gluons, the QGP was found to be strongly interacting~\cite{Gyulassy:2004zy,Muller:2006ee}. One of the evidence for this finding is the large anisotropy in final-state particle momentum distribution, called anisotropic flow, observed in non-head-on collisions~\cite{STAR:2005gfr,PHENIX:2004vcz,Back:2004je,Roland:2014jsa}. In those collisions, the nuclei are off center from each other, and the overlap portions of the nuclei on each other’s path form a region of an almond shape with extremely high temperature and energy density. The almond-shape region expands anisotropically because of interactions among the constituents, converting the initial spatial anisotropy into final-state momentum anisotropy~\cite{Ollitrault:1992bk,Heinz:2013th}. The interactions must be intense, close to the hydrodynamic limit with minimal viscosity to entropy density ratio ($\eta⁄s$)~\cite{Kovtun:2004de}, in order to attain the observed large anisotropy. In fact, the $\eta⁄s$ of the QGP was estimated to be $0.1$ in the unit of $\hbar$/$k_{B}$ ($\hbar$ is the reduced Planck constant and $k_{B}$ is the Boltzmann constant), close to the conjectured quantum limit of $1/4\pi$ by string theory~\cite{Kovtun:2004de}. In other words, the QGP observed in heavy-ion collisions at high energy behaves like a nearly perfect fluid~\cite{Gyulassy:2004zy}. 

More recently, large anisotropies have also been observed in small systems, like proton-proton and proton-nucleus collisions, where the interactions were not expected to be intense~\cite{Khachatryan:2010gv,CMS:2012qk,Abelev:2012ola,Aad:2012gla,Dusling:2015gta,Nagle:2018nvi}. This prompted several authors~\cite{He:2015hfa,PhysRevLett.120.012301,Kurkela:2018qeb} to suggest that intense interactions may not be a necessary prerequisite for anisotropy generation. By intense interactions we mean that each particle (constituent) suffers on average many collisions with other particles in the system. It was suggested that even a single collision may already be sufficient to cause enough anisotropic escape of particles resulting in a significant momentum anisotropy~\cite{Kurkela:2018qeb}. This casts doubt on the robustness of the nearly perfect fluid conclusion and begs the question how anisotropy builds up as a function of the intensity of interactions. This is, unfortunately, a difficult question to address with nuclear experiments beyond changing their beam species and collision impact parameter.

One can, however, tune interaction strengths in cold atom systems exploiting Feshbach resonances~\cite{RevModPhys.82.1225,PhysRevA.98.011601}. Large anisotropic expansion has been observed in strongly interacting cold Fermi gas~\cite{Ohara2002Science}, and such gases have been shown to behave like a nearly perfect fluid with $\eta⁄s$ estimated to be approximately $0.3$~\cite{ThomasPT2010,Schafer:2009dj}. 
A recent experiment observed generation of elliptic flow in  a system of few fermionic atoms~\cite{Brandstetter:2023jsy}.
These findings suggest that cold atom gases and the QGP may share some commonalities in the limit of intense interactions. With tunability of the interaction strength, one may examine not only the intense interaction regime but also the regimes of weak and intermediate interaction strengths. Together with manufacturability of the geometry, cold atom systems may offer a viable means to emulate the full range of interaction intensity in nuclear collisions.

In this Letter, we perform a cold atom expansion experiment, with systematic tuning of the interaction strength from zero to maximum and with two initial geometries of the gas cloud. Expansion anisotropies are measured as a function of the interaction strength. Quantitative comparisons are made to results from relativistic heavy-ion collisions. We observe a universality in the expansion anisotropy as a function of the number of collisions (termed opacity) from cold atoms to the hot QGP.

\section*{RESULTS}

\subsection*{Expansion of cold atom gases}
The experimental setup and methods for preparing and detecting cold atom systems are described in Supplemental Information. 
A cold Fermi gas of neutral $^{6}$Li atoms is prepared in the two lowest energy states with opposite spins. The gas is trapped in an optical dipole trap (ODT)~\cite{ODTGrimm} formed by crossing laser beams and evaporatively cooled down by lowering the trap depth. 
The beam crossing angle determines the shape of the trapped atom gas cloud to have an almond or cigar shape with its long axis in the axial ($z$) direction and the short axis in the transverse radial ($x$ and $y$) direction. 
The crossing angle is nominally $10^\circ$, and the anisotropic trapping frequencies are calculated to be $\omega_{r}/2\pi=2137$~Hz and $\omega_{z}/2\pi=187$~Hz along the radial and axial directions, respectively. The aspect ratio of the confined atom gas is given by $\beta=\omega_{z}/\omega_{r}=1/11.4$. 

The interaction strength (between atoms of opposite spins) of the cold atom system is tuned over a wide range with a homogeneous external magnetic field $B$. During evaporative cooling process, $B$ is set to $841$~G, near the Feshbach resonance point ($834$~G) to maximize inter-atomic interaction for efficient cooling~\cite{RevModPhys.82.1225}. 
At the end of this process, the ODT depth is stabilized; 
the magnetic field is then ramped down to the desired value $B$, and the ODT trap is turned off abruptly (in less than $1~\mu$s) to let the gas expand. 

The standard resonant absorption imaging (RAI) technique~\cite{BEC1995} is employed to evaluate the properties of expanding Fermi gas. 
The population in each spin state is observed to be $N\approx5.2\times10^{5}$, and the quantum degeneracy parameter is found to be $T/T_{F}=0.72$. Here, the gas temperature $T=4.6~\mu$K is determined by observing the time-of-flight (TOF) of interaction-free ballistic expansion at $B=527$~G; see Supplemental Information. The Fermi temperature is given by $T_{F}=(6\omega_{r}^{2}\omega_{z}N)^{1/3}\hbar/k_{B}=6.4~\mu$K. Since the degeneracy parameter is higher than the quantum degeneracy criterion ($T/T_{F}=0.5$)~\cite{PhysRevA.55.4346,PhysRevA.58.R4267}, the prepared cold atom Fermi gas is in normal phase and can be treated as a thermal cloud.

\begin{figure}[hbt]
    \centering
	\includegraphics[width=0.8\linewidth]{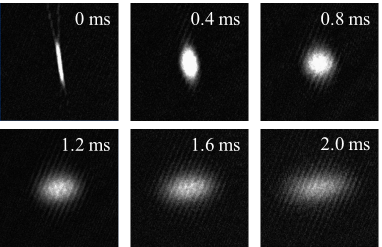}
	\caption{\textbf{Absorption images of the strongly interacting Fermi gas}\newline
    The images are taken with $B=831$~G at several time instances (indicated at top right of each image) after the gas is released from an anisotropic potential trap of aspect ratio of $\beta = 1/11.4$ (crossing angle $10^\circ$). The field of view of each image is $0.94$~mm~$\times~0.94$~mm.}
	\label{fig:fig1}
\end{figure}

When $B$ is tuned to the vicinity of the Feshbach resonance point, the gas expands anisotropically. The gas is imaged with the RAI technique at a given time. Since the imaging process is destructive, the gas is prepared with the same condition and the expansion experiment is repeated and imaged at each of several time instances. Figure~\ref{fig:fig1} shows the $z$-$x$ projective absorption images of the expanding gas with $B=831$~G at several time instances. The trapped gas is initially of cigar shape; once released, it expands and reverses its aspect ratio at $\sim 0.8$~ms; the aspect ratio reaches $\beta\approx1.95$ at $2.0$~ms. 

\begin{figure}[hbt]
    \centering
	\includegraphics[width=\linewidth]{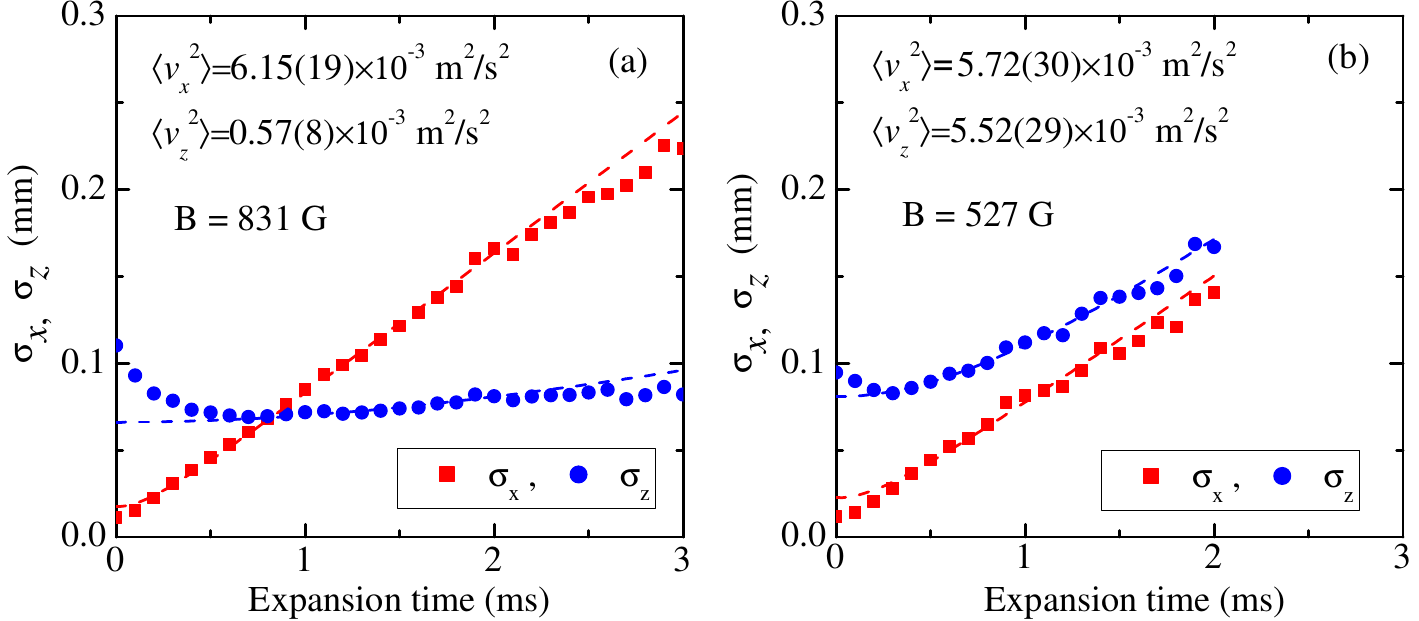}
	\caption{\textbf{RMS size of the cold atom gas}\newline
    The RMS sizes in the $z$ (axial) and $x$ (transverse) directions are plotted as functions of the expansion time. The external magnetic fields are $B = 831$~G (a) and $527$~G (b), corresponding to Feshbach resonance unitary limit and vanishing interaction, respectively. The dashed curves in (a) and (b) represent fits to Eq.~\ref{eq:(1)} within the ranges $1.0$--$2.0$~ms and $0.3$--$1.3$~ms, respectively; fitting parameters are indicated on the plots.}
	\label{fig:fig2}
\end{figure}

The $z$ and $x$ projections of the absorption images are fitted with Gaussian functions to extract the axial ($\sigma_{z}$) and transverse ($\sigma_{x}$) root-mean-square (RMS) size of the expanding Fermi gas, respectively. 
Figure~\ref{fig:fig2}(a) shows $\sigma_{z}$ and $\sigma_{x}$ as functions of the expansion time for $B=831$~G. 
The relative statistical uncertainty on each data point is approximately 1\%. 
Jitters in the data points reflect the size of systematic uncertainties, which become relatively large at long expansion times because of lower signal-to-noise ratios. The expansion in the axial direction is slow while it is rapid in the transverse (radial) direction; this causes the reversion of the aspect ratio. For comparison, the result with $B=527$~G is shown in Fig.~\ref{fig:fig2}(b) where the interaction vanishes and the expansion is ballistic; the aspect ratio does not reverse. Note that the measured axial size of the gas decreases initially because of absorption saturation; those axial data points are not used in subsequent analysis.

\subsection*{Expansion anisotropy}
The anisotropic expansion is caused by redistribution of momentum among particles. It happens at the initial stage of the expansion when interactions are intense. When the gas becomes dilute and interactions become negligible, the atoms stream freely. The late-time expansion of the cold gas cloud is treated as ballistic and can be described by~\cite{JOSA1989Molasses,Menotti2002}, 
\begin{equation}
	\sigma_{i}^{2}(t)=\sigma_{i}^{2}(0)+\left\langle v_{i}^{2}\right\rangle t^{2}
	\label{eq:(1)}\,,
\end{equation}
where $i = x,y,z$ are Cartesian components.
Here, $\sigma_{i}(0)$ represents the effective initial RMS size of the gas cloud (which differs from the initial size because of interactions during the early state of expansion), and $\left\langle v_{i}^{2}\right\rangle$ denotes the average squared velocity along the $i$-direction at long expansion time.

Practically, free streaming starts to set in when the mean free path $\lambda =\left(\langle \rho(t)\rangle\sigma_{s}\right)^{-1}$ becomes larger than the average size of the expanding gas~\cite{Giorgini2008,PhysRevLett.91.020402}, where $\langle\rho(t)\rangle$ is the mean atom number density at expansion time $t$, and $\sigma_{s}$ is the $s$-wave scattering cross section. At small $B$, this happens early, while for $B$ close to the Feshbach resonance point, it happens late. For $B=831$~G, the time is found to be $0.8$~ms, which is safely beyond the region of absorption saturation mentioned before. On the other hand, at long expansion times (after $\sim 2$~ms), the gas becomes increasingly dilute and the signal-to-noise ratio of the absorption image diminishes, resulting in progressively smaller RMS sizes at subsequent expansion times.
We therefore fit the $831$~G data to Eq.~\ref{eq:(1)} within expansion time $1.0$--$2.0$ ms. The fitted $\left\langle v_{i}^{2}\right\rangle$ values along $z$ and $x$ directions are written on the plots as shown in Fig.~\ref{fig:fig2}(a).
The $\left\langle v_{x}^{2}\right\rangle$ is significantly larger than $\left\langle v_{z}^{2}\right\rangle$, indicating stronger expansion in the $x$ direction than in the $z$ direction as a result of intense interactions among the atoms. Similarly, we fit the $527$~G data within $0.3$--$1.3$~ms as shown in Fig.~\ref{fig:fig2}(b). The fitted values $\left\langle v_{x}^{2}\right\rangle$ and $\left\langle v_{z}^{2}\right\rangle$ are comparable, indicating an isotropic and unchanged momentum distribution throughout the ballistic expansion process. 

The momentum anisotropy of the expanding gas can be quantified by the $v_{2}$ parameter~\cite{Ollitrault:1992bk}, 
\begin{equation}
	v_{2} = \frac{\left\langle v_{x}^{2}\right\rangle - \left\langle v_{z}^{2}\right\rangle}{\left\langle v_{x}^{2}\right\rangle + \left\langle v_{z}^{2}\right\rangle}
	\label{eq:(2)}\,.
\end{equation}
The uncertainty evaluation is described in Supplemental Information. 
Figure~\ref{fig:fig3} shows the $v_2$ parameter as a function of magnetic field $B$.
The magnetic field strength  determines the $s$-wave scattering length of the two-component cold Fermi gas, thereby the interaction strength.
The $v_{2}$ for $B=527$~G is practically zero, indicating an extremely weak interaction strength.
After an initial slow rise, the $v_{2}$ parameter increases rapidly with $B$, because of a rapid increase in the interaction strength (see Supplemental Information). 
Around the Feshbach resonant point at $B~=~834$~G, the $v_{2}$ value is large, 
well above $0.5$, and appears to saturate suggesting an unitarily limited cross section.
\begin{figure}[hbt]
    \centering
	\includegraphics[width=0.5\textwidth]{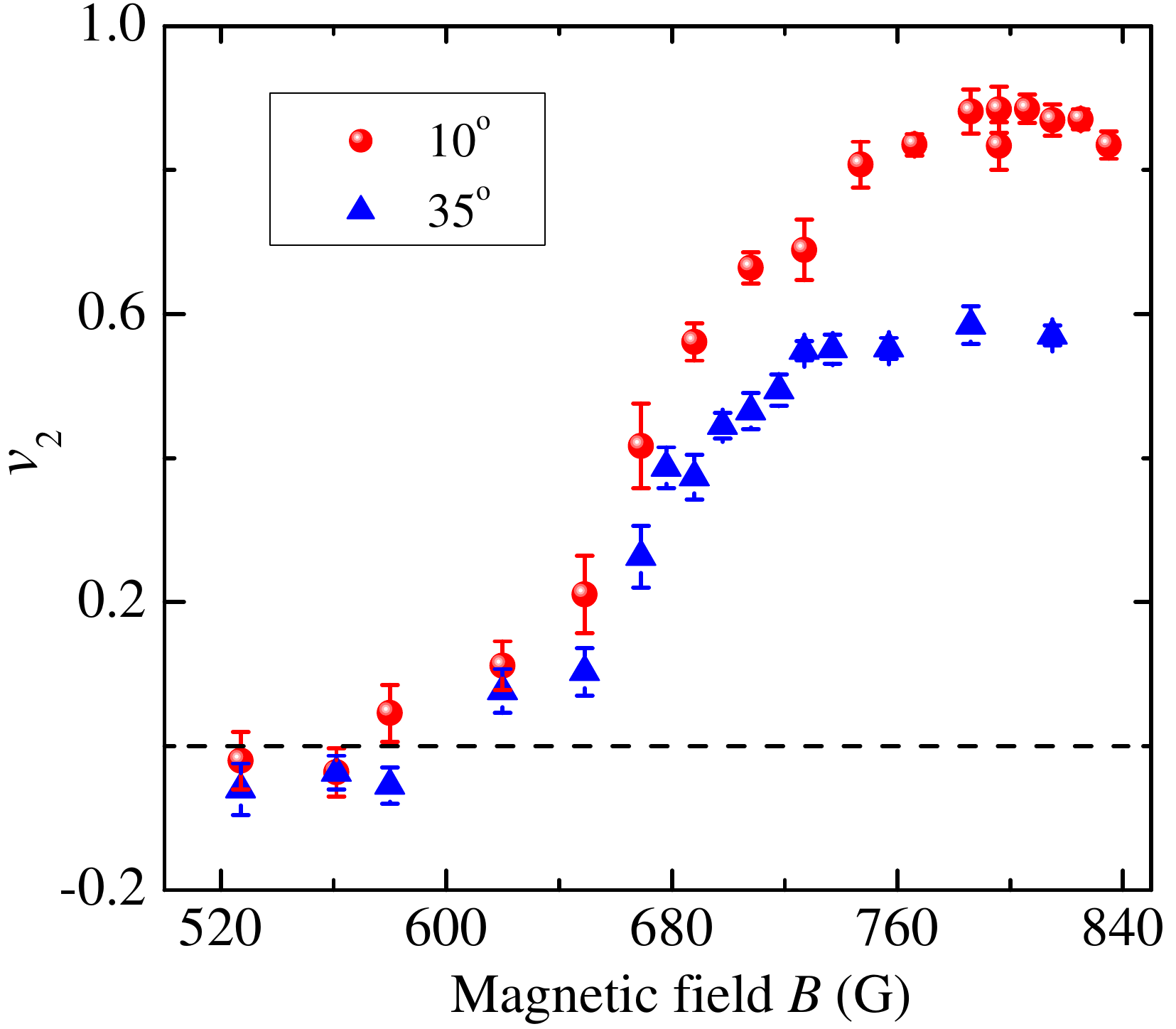}
	\caption{\textbf{The $v_{2}$ parameter as a function of the magnetic field $B$}\newline
    The $v_{2}$ parameter from the expansion of cold atom gases is shown as a function of the applied magnetic field $B$. The cold atom gases are released from two ODT potentials of differing shapes. One features a beam crossing angle of $10^\circ$ (red spheres), while the other has an angle of $35^\circ$ (blue triangles). 
    The error bars are statistical uncertainties of one standard deviation.}
\label{fig:fig3}
\end{figure}

In addition to the interaction strength, the parameter $v_{2}$ also depends on the magnitude of the initial spatial anisotropy of the trapped Fermi gas,
characterized by the eccentricity, 
\begin{equation}
    \varepsilon_{2} \equiv \frac{1-\beta^{2}}{1+\beta^{2}}\,.
\end{equation}
We thus have also build a second version of ODT with the laser beam crossing angle changed to $35^\circ$, so that the aspect ratio of trapped atom gas is reduced to $\beta = 1/3.17$. The prepared Fermi gas system has a total atom number $\sim 2.4\times 10^{5}$ in each spin state and a degeneracy parameter $T/T_{F} = 0.69$. 
The experiment is repeated, 
and the resultant $v_{2}$ is shown in Fig.~\ref{fig:fig3} as the blue filled triangles. The $v_{2}$ parameter is found to be smaller than those obtained with the $10^\circ$ crossing angle (red filled circles). 
The eccentricities of the two Fermi gases are $\varepsilon_{2} = 0.98$ and $0.82$, respectively. 
The observed difference in elliptic flow $v_2$ arises mostly from variations in the initial-state eccentricity $\varepsilon_2$. 
To account for this, $v_2$ is typically normalized by $\varepsilon_2$ when presenting data, as $\varepsilon_2$ fundamentally drives the final-state $v_2$. Additionally, differences in $v_2$ are influenced by slight variations in the interaction strengths of the gases at different crossing angles.
The ratio $v_2/\varepsilon_2$ is often referred to as the elliptic response coefficient, and in hydrodynamics  characterizes the amount of flow generated by a pressure gradient inside a liquid.
\subsection*{Opacity}
The amount of interaction can be quantified by the average number of collisions ($\ncoll$) an atom of the Fermi gas encounters during expansion, called opacity in the following. For a particle traversing a static medium of uniform density $\rho$ over distance $L$ with interaction cross section $\sigma$, it is simply $\ncoll=\rho\sigma L$. We estimate the opacity of our cold atom experiment by the average number of collisions a test atom originated at the center of the trapped gas cloud would encounter as it traverses outward as if the gas cloud is static. 
For initial Gaussian RMS radii of $\sigma_{z}$ and $\sigma_{x}$ (and $\beta = \sigma_{x}/\sigma_{z}$), it is given by
\begin{equation}
    \ncoll = \frac{\sigma_{s}}{4\pi}\oint{\rho_{0}\exp \left(-\frac{r^2\sin^{2}\theta}{2\sigma_{x}^2}-\frac{r^2\cos^{2}\theta}{2\sigma_{z}^2}\right)\sin\theta dr d\theta d\phi}
    = \frac{N\sigma_{s}}{4\pi\sigma_{z}^2}\frac{\arctan(\sqrt{1-\beta^{2}} / \beta)}{\beta\sqrt{1-\beta^{2}}}\,,
    \label{eq:n}
\end{equation}
where $\rho_{0}=\frac{N}{(2\pi)^{3/2}\sigma_{x}^{2}\sigma_{z}}$ is the center density of the cold gas cloud of a {\em single} spin. 

\begin{figure}[hbt]
    \centering
	\includegraphics[width=\linewidth]{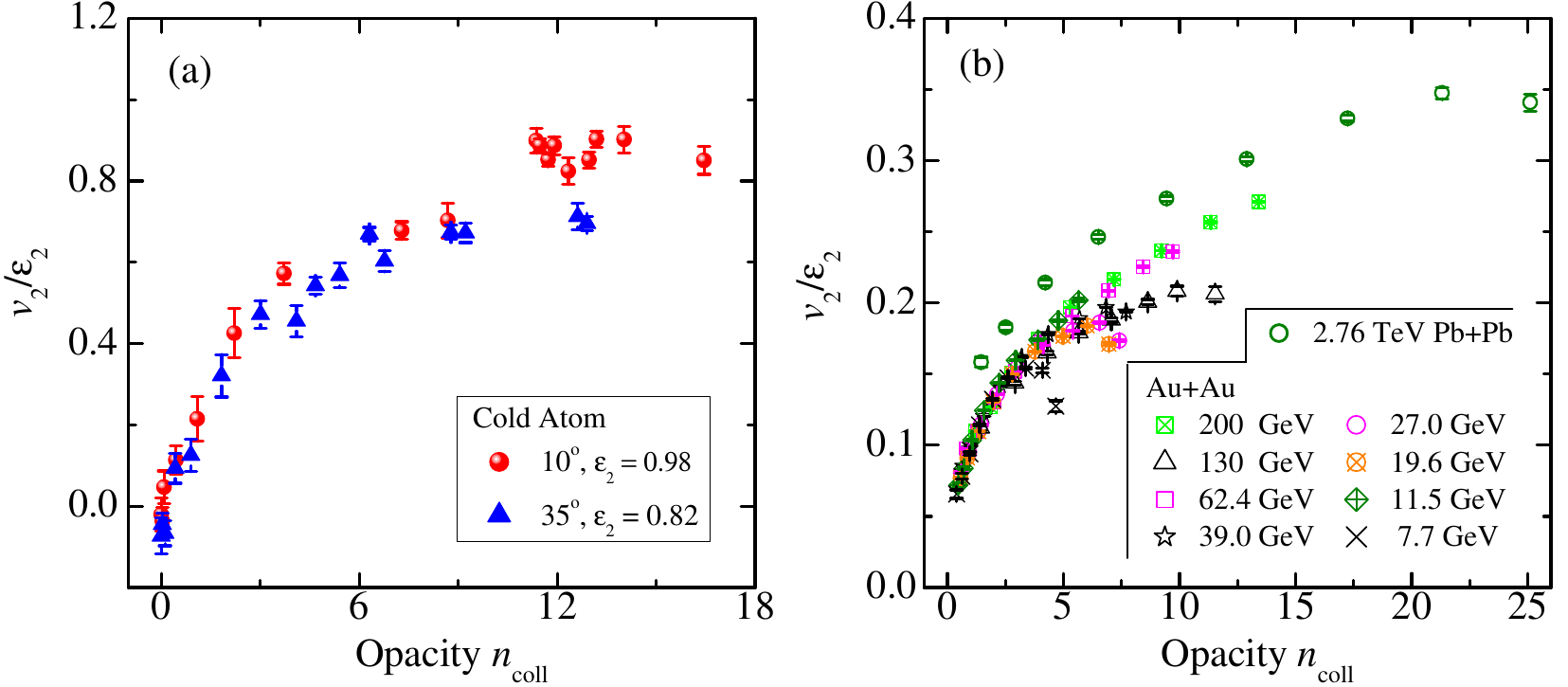}
	\caption{\textbf{Eccentricity normalized $v_{2}/\varepsilon_{2}$ as a function of opacity ($\ncoll$)}\newline
    The eccentricity normalized $v_{2}/\varepsilon_{2}$ is shown as a function of opacity ($\ncoll$) in cold atom gases obtained by this work (a) and in relativistic heavy-ion collisions (b). 
    The error bars are statistical uncertainties of one standard deviation.}
	\label{fig:fig4}
\end{figure}

Figure~\ref{fig:fig4}(a) shows $v_{2}$ (now divided by $\varepsilon_{2}$) as a function of $\ncoll$. The $v_{2}/\varepsilon_{2}$ versus $\ncoll$ data points appear to fall onto a common curve. Note that the $\ncoll$ values differ somewhat between the two cases because the initial geometries are slightly different between the two crossing angles; the smaller $v_{2}$ for the $35^\circ$ crossing angle is a combined effect of the smaller values of both $\varepsilon_{2}$ and $\ncoll$. 
The uncertainties on $\ncoll$ are described in Methods. 
These uncertainties are correlated between the two data sets at $10^\circ$ and $35^\circ$ cross angles, so the common trend seen in Fig.~\ref{fig:fig4}(a) is robust.

As mentioned in the introduction, strong elliptic flow has been observed in relativistic heavy-ion collisions. While the cold atom gas is three-dimensional, the heavy-ion collision system is effectively two-dimensional because the longitudinal beam direction is approximately Lorentz boost invariant. 
The anisotropy is on the transverse plane between $x$ and $y$ directions in heavy-ion collisions.
In interacting cold atom gases, the $z$ direction expands  little (see Fig.~\ref{fig:fig2}(a)), so the expansion is effectively also in two dimensions, albeit isotropically in the transverse direction. 
The anisotropy of the cold atom system is between the $x$ and $z$ directions; the $z$ direction is analogous to the $y$ direction in heavy-ion collisions but is stalled in expansion. %

The fireball created in heavy-ion collisions can be assumed to equilibrate at the typical strong interaction proper time $\tau\sim 1$~fm/$c$ with a longitudinal extent of $c\tau\sim 1$~fm, where $c$ is the speed of light in vacuum. The initial Bjorken density~\cite{Bjorken1983} of partons (quarks and gluons) can be estimated from the rapidity density of final-state particle multiplicity, $dN/dy$, by $\frac{dN/dy}{S_\perp\cdot c\tau}$ where $S_\perp$ is the transverse overlap area of the two colliding nuclei. Again, taking a test particle flying out from the center of the fireball, the opacity of the partonic matter created in heavy-ion collisions can be estimated as 
\begin{equation}
    \ncoll=\frac{\sigma}{c\tau}\frac{dN_{\rm ch}/dy}{\sqrt{\pi S_\perp}}\,,
    \label{eq:n_hi}
\end{equation}
where we use $\sigma=3$~mb as the parton-parton interaction cross section. Here, we have assumed an isentropic evolution with entropy conservation, so one gluon turns into one final-state pion, and we have simply used the charged hadron multiplicity density $dN_{\rm ch}/dy$. 
We note that while this choice of parameters is a common one it comes with significant uncertainties; this will be discussed further below.

Figure~\ref{fig:fig4}(b) shows $v_{2}/\varepsilon_{2}$ as a function of $\ncoll$ in gold-gold (Au+Au) collisions at RHIC~\cite{StarPRC2012,Star130GeV,Star39GeV,StarPRC2017} and lead-lead (Pb+Pb) colllisions at the LHC~\cite{ALICE:2010suc}. The data spans over a wide range in centralities or impact parameters ($b$) and at RHIC over a wide range in nucleon-nucleon center-of-mass energies ($\sqrt{s_{_{NN}}}$). 
The $v_2$ are measured by heavy-ion collision experiments, and the $\varepsilon_2$ is calculated by the Glauber model of the collision geometry; see Methods for details. 
As seen in Fig.~\ref{fig:fig4}(b) the different data sets scatter tightly around a common curve similar to the cold atom gas experiment. 
Also here the opacity estimation method is identical for all heavy-ion data, so the trend seen for the different energies in Fig.~\ref{fig:fig4}(b) is robust against common scale differences or systematic errors.
We note that it is not the first time the $v_2$ data are plotted as a function of scaling variables similar to the opacity used here; several authors including one of us have investigated the physics of anisotropic flow using scaling variables in heavy-ion collisions~\cite{Voloshin:1999gs,PHENIX:2013ktj,Lacey:2013qua,He:2015hfa,Kurkela:2018qeb,Ambrus:2024eqa}.
We also note that some deviations from the common trend are noticeable in Fig.~\ref{fig:fig4}(b) for the most-central heavy-ion collision data (large opacity, high multiplicity), particularly at lower energies. 
Such a deviation is also seen in the early results of the LHC data~\cite{CMS:2012zex}.
This is likely due to difficulties in modelling the initial shape of the nuclear overlap region in central collisions which is dominated by small fluctuations around an increasingly symmetric ``round'' average shape and is therefore extra sensitive to details of the geometry model used. This study focuses on the more general relationship between initial-state geometry and final-state momentum anisotropies and our conclusion are therefore not affected by the model dependence of initial state eccentricity calculations in the fluctuation dominated regime.

\subsection*{Universality in anisotropic expansion}
As seen in Fig.~\ref{fig:fig4}, while the opacities are comparable, the magnitude of $v_{2}/\varepsilon_{2}$ is larger in the cold atom gas than that in heavy-ion collisions.
There are at least two important differences between the observed $v_{2}$ from our cold atom experiment and the heavy-ion data. The first is technical: in heavy-ion experiment, the momentum is relativistic and measured particle by particle, and the $v_{2}$ parameter is averaged over all detected particles as 
$v_{2}=\left\langle \cos 2\phi \right\rangle=\left\langle\frac{p_{x}^{2}-p_{y}^{2}}{p_{x}^{2}+p_{y}^{2}} \right\rangle$, 
where $p_{x}$ and $p_{y}$ are the particle momentum components on the transverse $x$-$y$ plane~\cite{Poskanzer1998}; in cold atom systems which are non-relativistic, the average squared velocities are extracted from the two-dimensional density distributions obtained through absorption imaging, and the $v_{2}$ parameter is calculated by Eq.~\ref{eq:(2)}. To use the same definition as for cold atoms, the heavy-ion $v_{2}$ should be weighted by $p_{\perp}^{2}\equiv p_{x}^{2}+p_{y}^{2}$. Since $v_{2}$ is approximately proportional to transverse momentum $p_{\perp}$ in heavy-ion collisions, $v_{2}\propto p_{\perp}$~\cite{Poskanzer1998,Heinz:2013th}, and the $p_{\perp}$ distributions are typically exponential, $dN\propto p_{\perp}e^{-b p_{\perp}}dp_{\perp}$ where $b$ is a parameter related to the effective temperature~\cite{PRC2009pp}, the $p_{\perp}^{2}$ weighting would amount to a factor of $2$ difference: $\left.\frac{\int p_{\perp}^{2}v_{2}dN}{\int p_{\perp}^{2}dN}\right/\frac{\int v_{2}dN}{\int dN}=2$. In other words, the $v_{2}$ in heavy-ion collisions would be a factor of $2$ larger if it was calculated in the same way as in the cold atom experiment. 

\begin{figure*}[hbt]
    \centering
    \includegraphics[width=\linewidth]{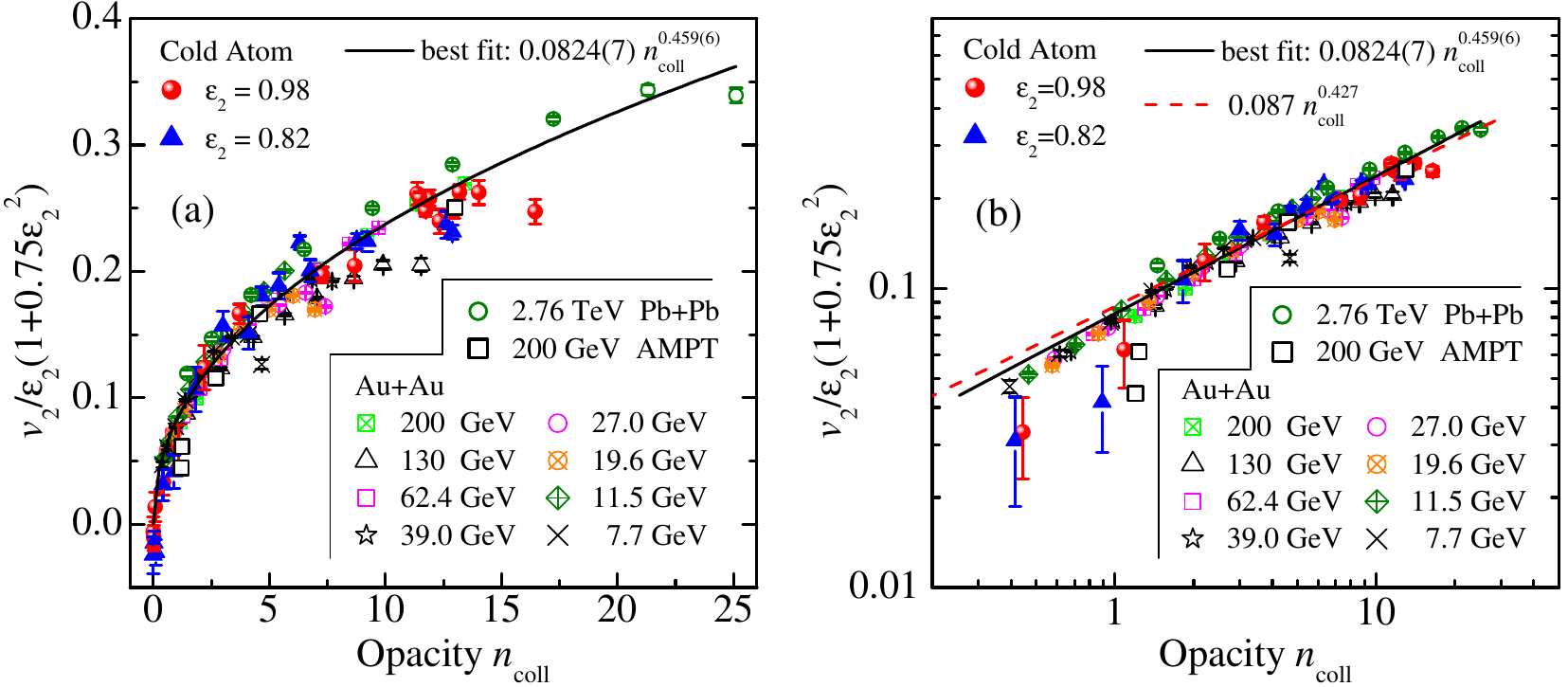}
	\caption{\textbf{Non-linearity corrected response $v_{2}/\left[\varepsilon_{2}(1+0.75\varepsilon_{2}^{2})\right]$ as a function of opacity ($\ncoll$)}\newline
    The non-linearity corrected response $v_{2}/\left[\varepsilon_{2}(1+0.75\varepsilon_{2}^{2})\right]$ is shown as a function of opacity ($\ncoll$) on linear scale (a) and log-log scale (b). The cold atom $v_{2}$ is first divided by $2$ because of the different means to measure the $v_{2}$ from that in heavy-ion collisions (see text). The solid line is a power-law fit to the combined data (statistically dominated by the heavy-ion data), and the dashed line in (b) is a separate power-law fit to the cold-atom data. The fit uncertainties quoted on the plots are statistical.}
	\label{fig:fig5}
\end{figure*}

The other difference is physical, namely non-linearity correction. The $\varepsilon_{2}$ of our cold atom gases is close to the maximum of unity, whereas that in heavy-ion collisions is in general smaller. 
According to hydrodynamic calculations of heavy-ion collisions~\cite{Noronha2016}, the relative cubic to linear response of $v_2$ to $\varepsilon_2$ increases with eccentricity from $0.3$ (central collisions, small eccentricities) to $0.6$ (midcentral to peripheral collisions, large eccentricities) with increasing $\varepsilon_2$, so the non-linearity effect can be significant at large $\varepsilon_2$. 
We varied the non-linearity coefficient between $0.5$ and $1$ and found that its precise value within this range is less critical (see Methods). 
In the following, we take its value to be $0.75$ (as the relevant eccentricity is rather large in peripheral heavy-ion collisions) and plot $v_{2}/\left[\varepsilon_{2}(1+0.75\varepsilon_{2}^{2})\right]$ in Fig.~\ref{fig:fig5}(a) in order to put systems of vastly different $\varepsilon_{2}$ onto the same footing. 
Both for the cold atom data and the heavy-ion data the scatter is significantly reduced and both data sets appear now to closely follow a single universal scaling curve.
Ideally, one would want to make the eccentricity of the potential trap comparable to those in heavy-ion collisions by increasing the crossing angle. However, this is not possible because of the space limitations around our experimental chamber. Note, for the two cold atom gases with different $\varepsilon_{2}$, the nonlinearity corrections are similar, differing by only $10\%$. As such, the cold atom gases appear consistent with each other even without nonlinearity corrections as in Fig.~\ref{fig:fig4}(a). It is noteworthy that the nonlinearity correction makes a noticeable difference for the LHC data from Fig.~\ref{fig:fig4}(b) to Fig.~\ref{fig:fig5}(a) because of the precise measurements. 
The nonlinearity correction improves the quantitative agreement between LHC and RHIC because the former receives a larger correction from the larger eccentricity (more peripheral collisions) at the same opacity value, and therefore reinforces the universal scaling seen in the data.

The estimation of $\ncoll$ for the cold atom system or for heavy-ion collisions is self-consistent individually and using a test particle originated at the system's center is common between the two cases. However, as we noted previously, the parameters we have chosen for heavy-ion collisions, although reasonable mainstream choices, have large uncertainties, such as the value used for the proper time. 
Consequently, the good and a priori unexpected quantitative agreement between the cold atom gas and the heavy-ion collisions could be accidental, by lucky choice of parameters.
We have therefore used a completely independent method to check the quantitative correspondence between opacity and flow response suggested by Fig.~\ref{fig:fig5}. The ``A Multi-Phase Transport'' (AMPT) {\em Monte Carlo} (MC) model~\cite{Lin:2004en} simulates parton-parton interactions in heavy-ion collisions. 
It has been tuned to describe flow in heavy-ion collisions to a good precision~\cite{Schukraft:2017nbn}.
As a microscopic transport model, both scale parameters, i.e.~flow response and opacity (the average number of collisions per parton), can be numerically extracted from the simulation and are shown in Fig.~\ref{fig:fig5}(a) as open squares for different model parameter combinations. 
The data point at the lowest opacity corresponds to 200~GeV deuteron-gold (d+Au) collisions ($b=0$~fm, $\sigma=3$~mb, $\ncoll=1.2$), the others to 200~GeV Au+Au collisions ($b=7.3$~fm corresponding to $\sim32\%$ centrality, $\sigma=0.6$, $1.5$, $3$, and $20$~mb, corresponding to $\ncoll=1.24$, $2.7$, $4.6$, and $13$, respectively). Note that the opacity is not strictly proportional to $\sigma$; this is because the number of collisions is ``unitarily" limited when cross section becomes large. The parton $v_2$ and $\ncoll$ values are taken from Ref.~\cite{He:2015hfa} except the $\sigma=0.6$ and $1.5$~mb points which are calculated by us following Ref.~\cite{He:2015hfa}; the eccentricity $\varepsilon_2$ are not presented in Ref.~\cite{He:2015hfa} so we have computed them by AMPT using the initial partons participating in the collision (see Methods). The $3$~mb AuAu simulation not only coincides with the experimentally measured values of $v_2/\varepsilon_2$, but also corresponds to an opacity similar to the one estimated for the corresponding data points ($\ncoll=4.57$ for 30–40\% centrality and $6.15$ for 20–30\% centrality). 
Likewise, the other AMPT results with different cross sections follow the general trends closely (even if d+Au is off by about a factor of 2), despite not corresponding to a simulation tuned to an actual experimental data point. The fact that the three vastly different systems, including a deliberately de-tuned MC simulation, cluster  tightly around a common universal scaling law, seems unlikely to be accidental. 

The common trend shown in Fig.~\ref{fig:fig5}(a) suggests that flow response to the initial geometry is dominantly controlled by the opacity $\ncoll$, no matter how the opacity is generated, whether by a cold atom Fermi gas, relativistic heavy-ion collisions, or in a rough transport model
of parton-parton interactions. The anisotropy rises sharply at small $\ncoll$, were the interactions are minimal, despite the fact that most particles escape without any interaction at all~\cite{He:2015hfa,Schukraft:2017nbn}. The rise flattens at large $\ncoll$, where the particles suffer many interactions and may behave like a liquid, asymptotically approaching ideal hydrodynamics~\cite{Gyulassy:2004zy,Ohara2002Science}. Since the opacity scales of AMPT and the cold atom gas are well constrained with small uncertainties, the universal trend can be used in turn to gauge the opacity in heavy-ion collisions which have inevitably many poorly constrained physical parameters.

Figure~\ref{fig:fig5}(b) shows the cold atom and heavy-ion data on a log-log plot. 
The response increases smoothly with increasing opacity, without any onset, saturation, or any other qualitative change apparent anywhere within the measured range. 
There are only quantitative changes; the flow starts at $\ncoll>0^+$ in a highly diluted system (average number of collisions per particle $\ncoll<1$) and evolves smoothly into the high-density hydrodynamic limit ($\ncoll\gg 1$).
The data points follow mostly a straight line on the log-log plot, indicative of a power-law behavior. Fitting the cold atom and heavy-ion data together to a power-law yields 
\begin{eqnarray}
   \frac{v_2}{\varepsilon_2(1+0.75\varepsilon_2^2)}=(0.082\mp0.006)\times\ncoll^{0.46\pm 0.03}\,,
\end{eqnarray}
where the quoted uncertainties are total uncertainties, dominated by systematic ones;
see Methods for assessment of the uncertainties. 
Separate fits to the cold atom data and the heavy-ion data yield an exponent of $0.43$ and $0.46$, respectively. 
Obviously the power-law rise cannot be asymptotically correct at $\ncoll\rightarrow\infty$ ($v_2$ is bound from above, see Eq.~\ref{eq:(2)}).
In the measured range of $\ncoll$, the power-law exponent is nearly equal to $1/2$, i.e.~a square-root behavior  indicative of random walks. The generation of anisotropy appears to be a result of statistical processes -- the interactions each constituent encounters are independent and add up incoherently; the flow response to the first interactions is strong whereas later interactions increase the flow response progressively less. 
The underlying microscopic physics of the system, on the other hand, seems to be of little importance. What matters is the opacity $\ncoll$, the average number of collisions each constituent suffers. 
A dilute system with strong interactions can have the same opacity as a dense system with weak interactions, or as a dense and strongly interacting system which is small or short lived; all of them would produce the same anisotropy.

It is interesting to note in Fig.~\ref{fig:fig5}(b) that globally the data continue to rise follow a straight line up to the densest systems measured at opacities as large as 25 in central Pb+Pb collisions at the LHC, showing no sign of saturation or approach to the hydrodynamic limit.
On the other hand, at small $\ncoll<1$ the data may be deviating from a single power-law behavior. It has been predicted that the $v_2/\varepsilon_2$ response becomes linear in opacity when approaching $\ncoll=0$~\red{\cite{Bhalerao:2005mm,Kurkela:2018qeb,Floerchinger:2021ygn,Ambrus:2024eqa}} (a  slope of $1$ on the log-log plot) and saturates at $\ncoll\gg 1$. The small $\ncoll$ behavior also seems to resemble the hydrodynamic attractor~\cite{Brewer:2019oha,Jankowski:2023fdz,Fujii:2024yce}. These aspects warrant further investigations with small-system collisions and cold atom gases in the weakly interacting and/or dilute regime, as well as with denser gases at higher opacities. As mentioned before, we used the traditional Glauber model to calculate the eccentricity of the initial overlap geometry. The initial condition of relativistic heavy-ion collisions is not fully settled, and other viable initial-state models, such as the IP-Glasma~\cite{Schenke:2012wb,Schenke:2012hg} and Trento~\cite{Moreland:2014oya} models, will likely alter the eccentricity values quantitatively. It will be interesting to investigate the implications of these models in the future. The observed universality in this work may, in turn, shed lights on the initial state of relativistic heavy-ion collisions.

\section*{DISCUSSION}

The common trend seen in Fig.~\ref{fig:fig5} suggests that the expansion dynamics is universal in interacting systems, from weak to strong, and dominantly governed by a single dimensionless scale parameter, the opacity.
This is remarkable considering the vast differences between the two systems -- the density of the QGP is $\sim 10^{39}$~cm$^{-3}$ and that of a typical cold atom gas is $\sim 10^{12}$~cm$^{-3}$ (about $7$ orders thinner than air); the temperature of the QGP is $\sim 10^{12}$~K and that of a cold atom gas is $\sim 10^{-6}$~K; the physics governing the QGP is the strong interaction of quantum chromodynamics (QCD) and that governing cold atom gases is electromagnetic interaction of quantum electrodynamics (QED). 

Strongly interacting systems are common in nature, e.g.~black holes~\cite{Hawking1975}, neutron stars~\cite{Hewish1968}, strongly coupled Bose fluid~\cite{LeClair2011}, superfluid liquid helium~\cite{Penrose1956}, and other condensed matter~\cite{Bloch2008RMP} and quantum systems~\cite{Georgescu2014RMP}, in addition to the QGP and cold atom Fermi gas. Studies of quantum gases, with their advantages of tunable interactions and variable geometries, may shed lights on non-perturbative many-body interactions in a variety of disciplines in the future~\cite{Zinner2013,Levinsen2017}.

Although we introduced new results and used new variables in our cold atom experiments, similar measurements have been performed before. Likewise, heavy-ion collisions have been extensively studied with various scaling parameters. The main advance and value of our work is bringing the two sets of vastly different systems together by the same opacity parameter to allow a quantitative comparison. 
As with any scaling behavior, the observed universality of anisotropic expansion is presumably of limited accuracy; it may not be sensitive to detailed properties of the matter under investigation (e.g.~viscosity, equation of state), which can be measured by more differential and detailed comparisons between data and specific models. Evaluation of the scaling accuracy by both theory and experiments should help to quantify its (in)sensitivity to specific matter properties.

Our cold atom experiment emulator can be improved in a number of ways. For example, the atom density may be increased to reach higher opacity to investigate the expansion behavior at even larger amount of interactions; 
it may alternatively be decreased to explore the dynamics of interacting systems similar to small-system collisions; 
and the geometry of the gas cloud can be varied and a triangular geometry may be manufactured to study triangular flow expansion~\cite{Alver:2010gr}. 
The temperature of the atom gas can be lowered to superfluid regime, where the dynamics are likely different, which may offer a means to study effects of phase transition on expansion dynamics.
The cold atom experiment can also be extended for other tests. For example, with addition of an ion trap, one may shoot an energetic ion through a cold atom gas to study their interactions. This would be similar to the jet quenching phenomenon~\cite{PhysRevLett.68.1480} observed in relativistic heavy-ion collisions, another evidence for the strongly interacting QGP besides the anisotropic flow. At ion speed higher than the speed of sound of the atom gas, the Mach-cone shock wave phenomenon may be studied.

In summary, we have carried out an experiment of cold $^6$Li atoms in normal gas phase with two trap geometries to systematically study the expansion behavior as a function of the interaction strength, tuned by an external magnetic field. The amount of interactions is characterized by the opacity variable $\ncoll$, effectively the number of collisions a test atom has suffered before free streaming. 
It is found that the aspect ratio of the cold atom cloud reverses already with minimal amount of interactions, indicating momentum (expansion) anisotropy. This expansion anisotropy builds up quickly at small $\ncoll$, without the need of a large amount of interactions, and is found to increase with $\ncoll$ more slowly at larger $\ncoll$.

The cold atom system is compared to heavy-ion collisions where strong elliptic flow has been observed suggesting a nearly perfect fluid formed in those collisions.
A universal behavior is quantitatively observed between these two vastly different systems, 
where the non-linearity corrected, eccentricity normalized elliptic anisotropy parameter $v_{2}$ follows a common trend in opacity $\ncoll$. 
The trend can be described by a power-law $\ncoll^{0.46\pm0.03}$, close to a $\sqrt{\ncoll}$ dependence suggesting random-walk behavior. 
This universality suggests that vastly different interacting systems can respond similarly in their expansion dynamics, regardless of the nature of the interactions, the interaction strength, or the system size, as long as the opacity $\ncoll$ value is the same.
The universality potentially unifies a variety of disciplines in nature, from the weakly interacting dilute systems of gases to the strongly interacting quark-gluon plasma of the early universe.


\clearpage

\section*{METHODS}

\subsection*{Relative uncertainty on opacity}\label{methods:opacity}
For thermal Fermi gas with $T>0.5~T_{F}$, the opacity $\ncoll$ is estimated by Eq.~\ref{eq:n} for the initial trapped cloud, where $\Phi\equiv\arctan(\sqrt{1-\beta^{2}}/\beta)/(\beta\sqrt{1-\beta^{2}})$ is fixed for a given ODT cross angle. The uncertainty on $\ncoll$ is primarily determined by those on $N$, $\sigma_s$, and $\sigma_z$. 

To determine $N$, the expansion time is fixed in our experiment to ensure that the observed absorption images have a peak optical density (OD) $\sim 2$ (see Eq.~(2) in Supplemental Information) 
to yield a high signal-to-noise ratio. 
This results in a relative atom number uncertainty of less than $5\%$. 
To reliably determine the initial RMS radius $\sigma_{z}$, we switch off the ODT to let the gas cloud expand for $0.3$~ms before detecting it with the RAI method. 
We then fit the 1D integrated axial OD profile by a Gaussian function. 
The relative statistical uncertainties on the cloud radii from fitting are typically $\sim 1\%$ (see examples illustrated in Fig.~S4 in Supplemental Information). 
However, the initial axial expansion is slow and saturation effect prolongs, the uncertainty on the estimated initial $\sigma_{z}$ is relatively significant, up to $\sim 6\%$. 
The uncertainty of the magnetic field $B$ at $10$~mG is determined using the RF spectrum technique, from which the relative uncertainty on the $s$-wave scattering length $a_{s}$ (see Supplemental Information) for each $B$ is estimated to be significantly less than $1\%$. 

For $|ka_{s}|\ll 1$ where $k$ is the typical relative wave number of the scattering atoms (see Supplemental Information), $n\approx N\Phi a_{s}^2/\sigma_{z}^2$, and the upper bound on $\Delta\ncoll/\ncoll$ is $+13\%$. For $|ka_{s}|\gg 1$, $n\approx N\Phi \hbar^{2}/(2m^{2}\omega_{z}^{2}\sigma_{z}^{4})$ and becomes independent of $a_{s}$. We estimate an upper bound on $\Delta\ncoll/\ncoll$ in the strongly interacting regime to be $+24\%$.

\subsection*{Heavy-ion collision data and AMPT simulations}\label{methods:hi}
The Au+Au collision data are taken from STAR measurements. 
For Au+Au collisions at $\sqrt{s_{_{NN}}}=200$ and $62.4$~GeV, the elliptic flow $v_{2}$ data are taken from Ref.~\cite{PhysRevLett.92.112301}, 
and the charged hadron multiplicity pseudorapidity density $dN_{\rm ch}/d\eta$ 
are taken from Ref.~\cite{PRC2009pp}. 
For Au+Au collisions at $\sqrt{s_{_{NN}}}=130$~GeV, the $v_{2}$ data 
are taken from Ref.~\cite{Star130GeV}, and the $dN_{\rm ch}/d\eta$ multiplicity data are taken from Ref.~\cite{PRC2009pp}. 
For Au+Au collisions at $\sqrt{s_{_{NN}}}=39$, $27$, $19.6$, $11.5$, $7.7$~GeV, the $v_{2}$ data 
are taken from Ref.~\cite{Star39GeV}, and the charged hadron multiplicity rapidity density $dN_{\rm ch}/dy$ are taken from Ref.~\cite{StarPRC2017}. 
For Pb+Pb collisions at $\sqrt{s_{_{NN}}}=2.76$~TeV, the $v_{2}$ data are taken from Ref.~\cite{ALICE:2010suc}, and the charged hadron multiplicity pseudorapidity density $dN_{\rm ch}/d\eta$ are taken from Ref.~\cite{ALICE:2010mlf}. 
The $v_2$ data were obtained from the two-particle cumulant method. 
All heavy-ion data were measured at midrapidity of approximately $2$ units of pseudorapidity.

For uniformity, the two-particle cumulant eccentricities $\varepsilon_2$ and overlap area $S_{\perp}$ are calculated from nuclear collision geometry using the Glauber model with nucleonic degrees of freedom~\cite{Miller:2007ri,Loizides:2017ack}. The results are qualitatively consistent with those calculated by similar means in the experimental publications~\cite{PRC2009pp,Star130GeV,Star39GeV}.

The systematic uncertainties are usually large in central (small impact parameter) collisions, which is seen in the spread of the data points in Fig.~\ref{fig:fig4}(b) and Fig.~\ref{fig:fig5}. 

For the AMPT eccentricities, the average $\varepsilon_2$ values are calculated using the so-called participant plane determined by the initial spatial distributions of partons, in the similar manner in which the $v_2$ was calculated in Ref.~\cite{He:2015hfa}. The AMPT results were calculated using all final partons participating in the collision.

\subsection*{Uncertainty estimate on the power-law behavior}\label{methods:powerlaw}
\begin{figure*}[hbt]
    \centering
	\includegraphics[width=\linewidth]{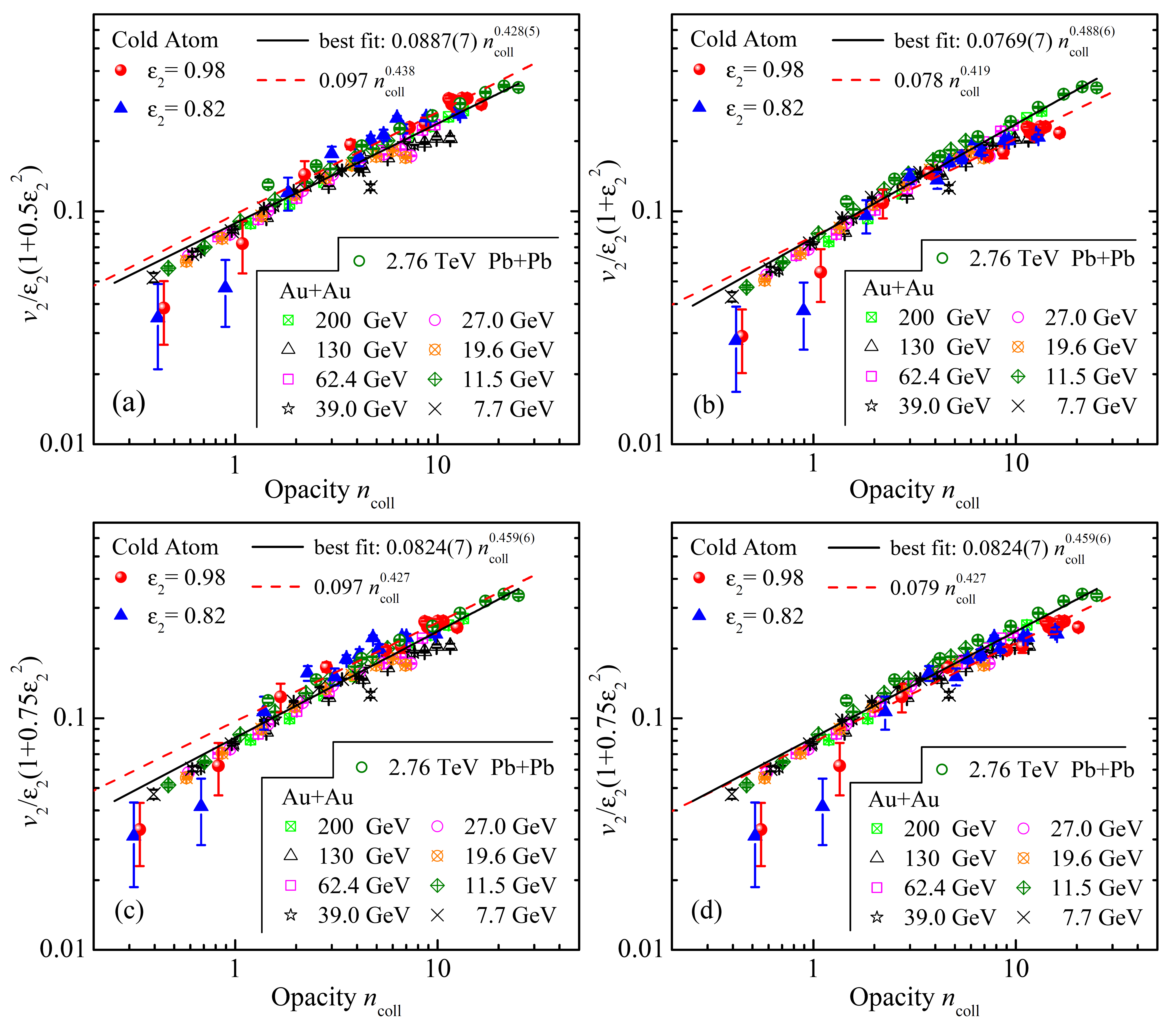}
	\caption{\textbf{Systematic uncertainty assessment on the power-law behavior}\newline
    Plots are similar to Fig.~\ref{fig:fig5}(b), but with nonlinearity coefficient 0.5 (a) and 1.0 (b), and with the opacity values of the cold atom data scaled by a factor of 0.76 (c) and 1.24 (d). The fit uncertainties quoted on the plots are statistical.}
	\label{fig:loglog4}
\end{figure*}

We have varied the non-linearity coefficient from the default $0.75$ to $0.5$ and $1.0$ and replotted the data in Fig.~\ref{fig:loglog4}(a) and (b), with the power-law fit parameters indicated on the plots. We have also shifted the opacity values for the cold atom gases by $\pm 24\%$ in both ways even though the systematic uncertainty in its estimates is only an one-sided upper bound; 
they are plotted in Fig.~\ref{fig:loglog4}(c) and (d). There are no qualitative visual changes.

We do not try to vary the opacity values in the heavy-ion data because they appear to be consistent with transport model calculations and because it is only the relative shift in the opacities between heavy-ion data and cold atom data that matters. 
Our opacity value is the average number of collisions a test particle originated at the system's center suffers while traversing a static medium. A more realistic number of collisions averaged over all particles in a dynamically expanding system may be computed by MC means. It presumably scales with our calculated opacity by a constant factor, without affecting the exponent of the power-law dependence.

The above variations provide reasonable assessments of the uncertainties in the power-law behavior. 
Based on the fitting results in Fig.~\ref{fig:loglog4}, we estimate the uncertainties on the power-law amplitude and exponent to be $\mp0.006$ and $\pm0.03$, respectively, and anticorrelated.

\bibliography{mainbib}
\clearpage

\section*{DATA AND CODE AVAILABILITY}
Experimental data and computer codes to process and analyze the experimental data are available from the authors upon request.

\section*{ACKNOWLEDGMENTS}
F.W.~thanks Dr.~Jurgen Schukraft for fruitful discussions. 
We thank Dr.~An Gu and Dr.~Ziwei Lin for help on the AMPT model. 
This work was supported by Huzhou University Educational and Research Fund, National Natural Science Foundation of China (12035006, 12075085, 12205095, 12275082), Ministry of Science and Technology of China (2020YFE020200).

\section*{AUTHOR CONTRIBUTIONS}
K.L.~designed and built the experimental set-up;
K.L.~and H.S.~carried out the experiment, took and analyzed the data;
K.L.~performed the theoretical calculations; 
H.X.~performed the AMPT and Glauber model calculations; 
H.X.~and Y.S.~advised on the theoretical modeling and simulations; 
K.L.~and F.W.~conceived the idea and wrote the manuscript with contributions from all authors.

\section*{DECLARATION OF INTERESTS}
The authors declare no competing interests.


\clearpage
\setcounter{figure}{0}
\renewcommand{\thefigure}{S\arabic{figure}}
\setcounter{table}{0}    
\renewcommand\thetable{S\arabic{table}}    

\section*{SUPPLEMENTAL METHODS}

\subsection*{The cold Fermi gas of $^{6}$Li atoms}\label{si:exp}
The laser source constructing the optical dipole trap (ODT)~\cite{ODTGrimm} is derived from the infrared emission of a fiber laser (IPG Photonics, YLR-100-1064-LP) which is centered at $1064$~nm and has a maximum power of $100$~W. The beam waist of the ODT laser is $\sim 40~\mu$m, corresponding to a calculated trap depth of $\sim 1.8$~mK at full power. The laser is split into two parts equally, which propagate on the horizontal plane with orthogonal polarizations and are finally focused and intersected at the focal points with a crossing angle $\theta$. The crossed beams produce an anisotropic trapping potential, which can be approximately described by a three-dimensional harmonic oscillator, $U=\frac{1}{2}m(\omega_{x}^{2}x^{2}+\omega_{y}^{2}y^{2}+\omega_{z}^{2}z^{2})$, where $m$ is the mass of $^{6}$Li atom, and $\omega_{i}$ is the trapping frequency along $i$-direction ($i=x,y,z$) and $\omega_{x}=\omega_{y}=\omega_{r}$. The trapping frequencies are determined mainly by the potential depth $U$, the beam waist $w_{0}$, and the crossing angle $\theta$. In our experiment, the trajectory of evaporative cooling is optimized by observing the stabilization of the atom gas temperature around one-tenth of the trap depth ($T \sim U/10$), confirming that our treatment is valid using the harmonic approximation for the optical potential.  

The ODT is loaded with a sample of cold $^{6}$Li atoms directly from a magneto-optical trap (MOT)~\cite{PhysRevA.46.4082}.  After loading, the laser power is ramped down from full power to $30$~W where the trap depth is $\sim 550~\mu$K. The initial population in each spin state of $^{6}$Li atoms, $|1/2,\pm 1/2\rangle$~\cite{PhysRevLett.85.2092}, is observed to be approximately $8.2\times 10^{5}$ using the standard RAI technique at high magnetic field, with a population difference between the two states less than $12\%$. 

The temperature of the atom gas is further lowered through two-stage evaporative cooling~\cite{PhysRevLett.88.120405,PhysRevLett.87.010404}. The first stage is the so-called free evaporative cooling process, during which the external magnetic field $B=841$~G is switched on and the laser power is maintained at $30$~W for $300$~ms. Then the second stage, a forced evaporative cooling process, is conducted by lowering the laser power according to an optimized ramping trajectory. The complete process of forced evaporative cooling typically takes 900~ms. Finally the ODT laser power is reduced to $1$~W per beam, corresponding to a trap depth of $\sim 50~\mu$K. The $s$-wave scattering length $a_{s}=-2.57~\mu$m at $B=841$~G. The interaction strength parameter $|k_{F}a_{s}| \gg 1$ holds for the entire evaporative cooling process. Here, $k_{F}=\sqrt{2m k_{B}T_{F}/\hbar^{2}}$ is the Fermi wave number~\cite{Giorgini2008}.

The properties of the prepared Fermi gases are tabulated in Table \ref{tab:gas}. The trap depths are based on theoretical calculations. The temperature $T$ is measured with TOF of ballistic expansion at $B=527$~G. 
The Fermi temperature, $T_{F}=(6\omega_{r}^{2}\omega_{z}N)^{1/3}\hbar/k_{B}$, is determined using the model of an ideal Fermi gas in a harmonic trapping potential, whereas $N$ represents the number of atoms in spin state $|1\rangle$ as depicted in Fig.~\ref{fig:figS1}(a). Note that $N$ is observed to be less than the initial loaded population due to the evaporation process that occurs during the cooling stages. 

\begin{table}[hbt]
	\caption{Properties of the prepared Fermi gases.
     $N$ is the total number of atoms in spin state $|1\rangle$, $T$ is the gas temperature, and $T_{F}$ is the Fermi temperature of the gas.}
	\label{tab:gas}
    \centering
	\begin{tabular}{c c c c c c}
		\hline
		Crossing angle & Trap depth & $N (\times 10^{5})$ & $T$ & $T_{F}$ & $T/T_{F}$\\
		\hline
		$10^\circ$ & $50~\mu$K & $5.2 \pm 0.3$ & $4.6~\mu$K & $6.4~\mu$K & $0.72$ \\
		$35^\circ$ & $57~\mu$K & $2.4 \pm 0.3$ & $6.4~\mu$K & $9.4~\mu$K & $0.69$ \\
		\hline
	\end{tabular}
\end{table}

\subsection*{Feshbach resonance}\label{si:feshbach}
\begin{figure}[hbt]
	\includegraphics[width=\linewidth]{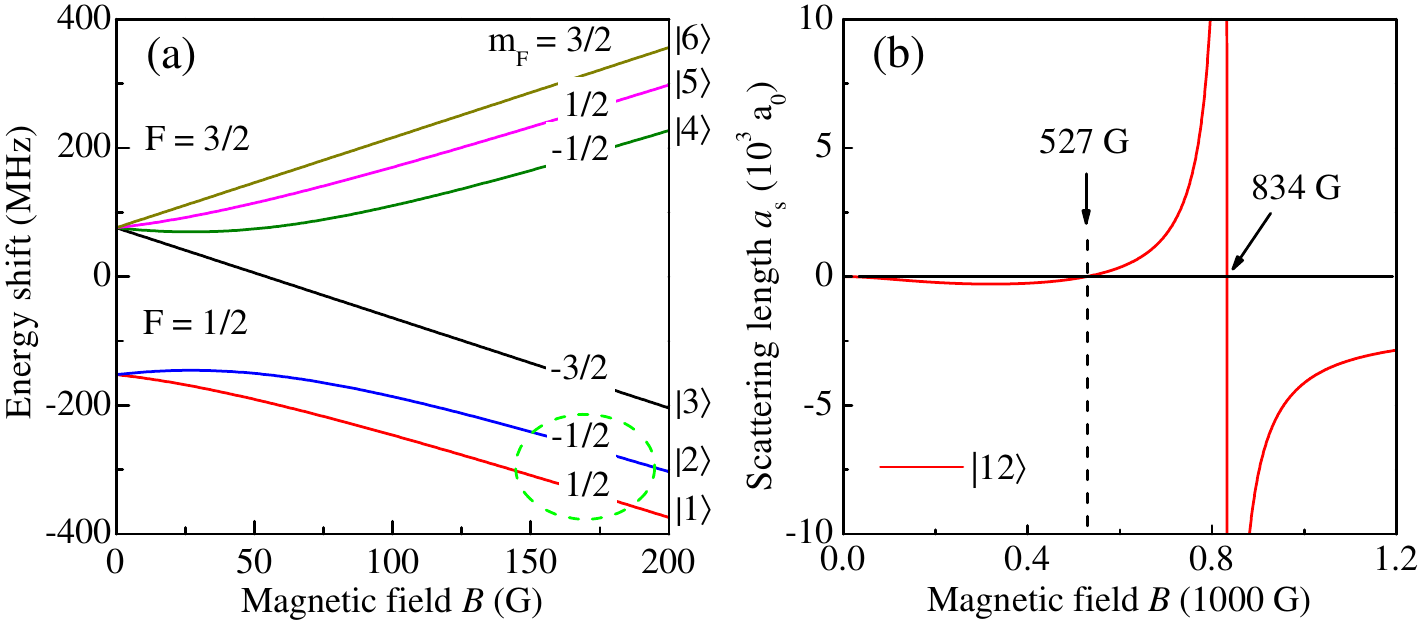}
	\caption{\textbf{$^{6}$Li atoms in magnetic field and Feshbach resonance}\newline
    (a) Energy splitting of the $^{6}$Li $2S_{1/2}$ ground state in external magnetic field. (b) Theoretical $s$-wave scattering length $a_{s}$ between states $|1\rangle$ and $|2\rangle$ as a function of the external magnetic field $B$; $a_{0}\approx 0.53 \times 10^{-10}$~m is the Bohr radius.}
	\label{fig:figS1}
\end{figure}

At zero magnetic field, the $2S_{1/2}$ ground state of $^{6}$Li atoms splits into two hyperfine states with total angular momentum ($\hbar⁄2$ and $3\hbar⁄2$). In an external magnetic field $B \neq 0$, they split further into six Zeeman states as shown in Fig.~\ref{fig:figS1}(a), labeled $|i\rangle$  ($i=1,2,...,6$)~\cite{PhysRevLett.85.2092}. The ODT can trap atoms in all of these spin states. For simplicity, our confined cold atoms are prepared in the two lowest energy states $|1\rangle$ and $|2\rangle$.

The interaction strength between $^{6}$Li atoms populated in spin states $|1\rangle$ and $|2\rangle$ can be described with a single parameter, the $s$-wave elastic scattering length $a_{s}$, which is controllable with the external magnetic field $B$ as shown in Fig.~\ref{fig:figS1}~\cite{PhysRevLett.89.273202}. The Feshbach resonant point between atom populated in spin states $|1\rangle$ and $|2\rangle$ is measured with Radio-Frequency (RF) Spectroscopy technique with high precision~\cite{PhysRevLett.94.103201,PhysRevLett.110.135301}. In our experiment, we employ the RF spectroscopy method to determine $B$ with an accuracy of 10 mG. The $s$-wave scattering cross section between atoms of different spin states is given by~\cite{,PhysRevLett.85.2092}
\begin{equation}
	\sigma_{s}=\frac{4\pi a_{s}^{2}}{1+k^{2} a_{s}^{2}}
	\label{eq:(B1)}\,,
\end{equation}
where $k=\sqrt{2m k_{B}T/\hbar^{2}}$ is the typical relative wave number of two colliding atoms. For $|k a_{s}| \gg 1$, the scattering cross section becomes unitary limited, $\sigma_{s}=4\pi/k^{2}$; for $|k a_{s}| \ll 1$, $\sigma_{s}=4\pi a_{s}^2$.

As shown in Fig.~\ref{fig:figS1}(b), $a_{s}$ is tuned with the same magnetic field $B$ that causes the splitting of hyperfine states. When the magnetic field is tuned across $527$~G, $a_{s}$ crosses zero and becomes positive. When $B$ is further approaching the Feshbach resonance point of $834$~G, $a_{s}$ grows to positive infinity. For several representative magnetic field strengths of $B$ = $685$, $763$ and $831$~G, the scattering lengths are $a_{s} = 0.07$, $0.236$, and $18.6~\mu$m, and the values of interaction strength $k_{F}a_{s}$ are $0.87$, $3.5$, and $241$, respectively.

\subsection*{Resonant absorption imaging (RAI)}\label{si:rai}
\begin{figure}
	\includegraphics[width=\linewidth]{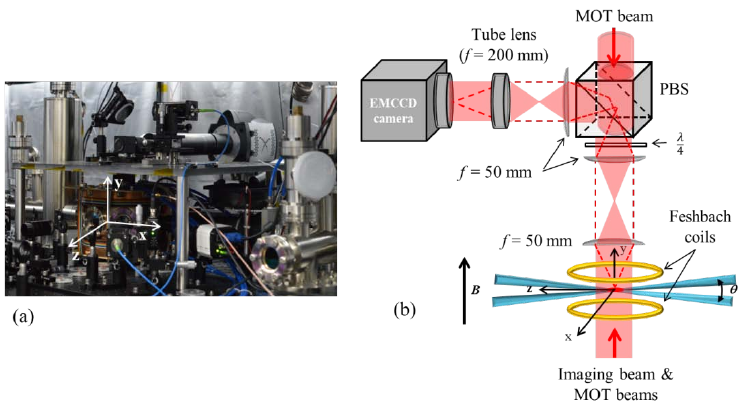}
	\caption{\textbf{Experiment setup}\newline
    (a) Setup of the Fermi gas expansion experiment. (b) Schematic drawing of the detection method by absorption imaging. The magnetic field $B$ generated by a pair of Holmholtz coils (Feshbach coils) is in vertical direction ($y$-axis). Axial direction lies in $z$-axis, and the horizontal radial direction is $x$-axis. The $\theta$ refers to the crossing angle.}
	\label{fig:figS2}
\end{figure}

In our experiment, $z$-$x$ plane is horizontal with the $z$-axis lies in the axial direction and the $x$-axis in one of the radial directions, as shown Fig.~\ref{fig:figS2}. The external magnetic field $B$ is along the vertical direction ($y$-axis, pointing upward). The imaging beam propagates in the vertical direction and is $\sigma^{-}$ polarized (antiparallel to $B$). During imaging process, only the ODT is switched off while $B$ is kept on. After expansion time $t$, a $10~\mu$s probe pulse is fired with light frequency finely tuned to be on resonant with the specific spin state at the given $B$. The behavior of the expanding Fermi gas can be assessed by imaging atoms populated in any one of the spin states ($|1\rangle$ and $|2\rangle$). In our experiment, we mainly detect atoms in spin state $|1\rangle$.

The center intensity of the probe light is $I_{\rm probe} = 0.23$ mW/cm$^{2}$, corresponding to a saturation parameter $I_{\rm probe}/I_{\rm sat}\sim 0.09$, which ensures the population in the spin state is not disturbed during imaging. Here, $I_{\rm sat} = 2.54$~mW/cm$^{2}$ is the saturation intensity of D2-line transition of $^{6}$Li atom. The imaging system consists of two stages. The first stage is a 1:1 image-relay, which is formed with an $f = 50$~mm lens pair. The image-relay has a numerical aperture of $0.2$, corresponding to an optical resolution $\sim 2~\mu$m. The relayed image is magnified and projected onto the electron multiplying charge coupled device (EMCCD) camera. The pixel size of the camera sensor is $13~\mu$m and the overall magnification of the imaging system is calibrated to be $\sim 4.15$, resulting in a spatial resolution of $3.1~\mu$m. 

\begin{figure}
	\includegraphics[width=\linewidth]{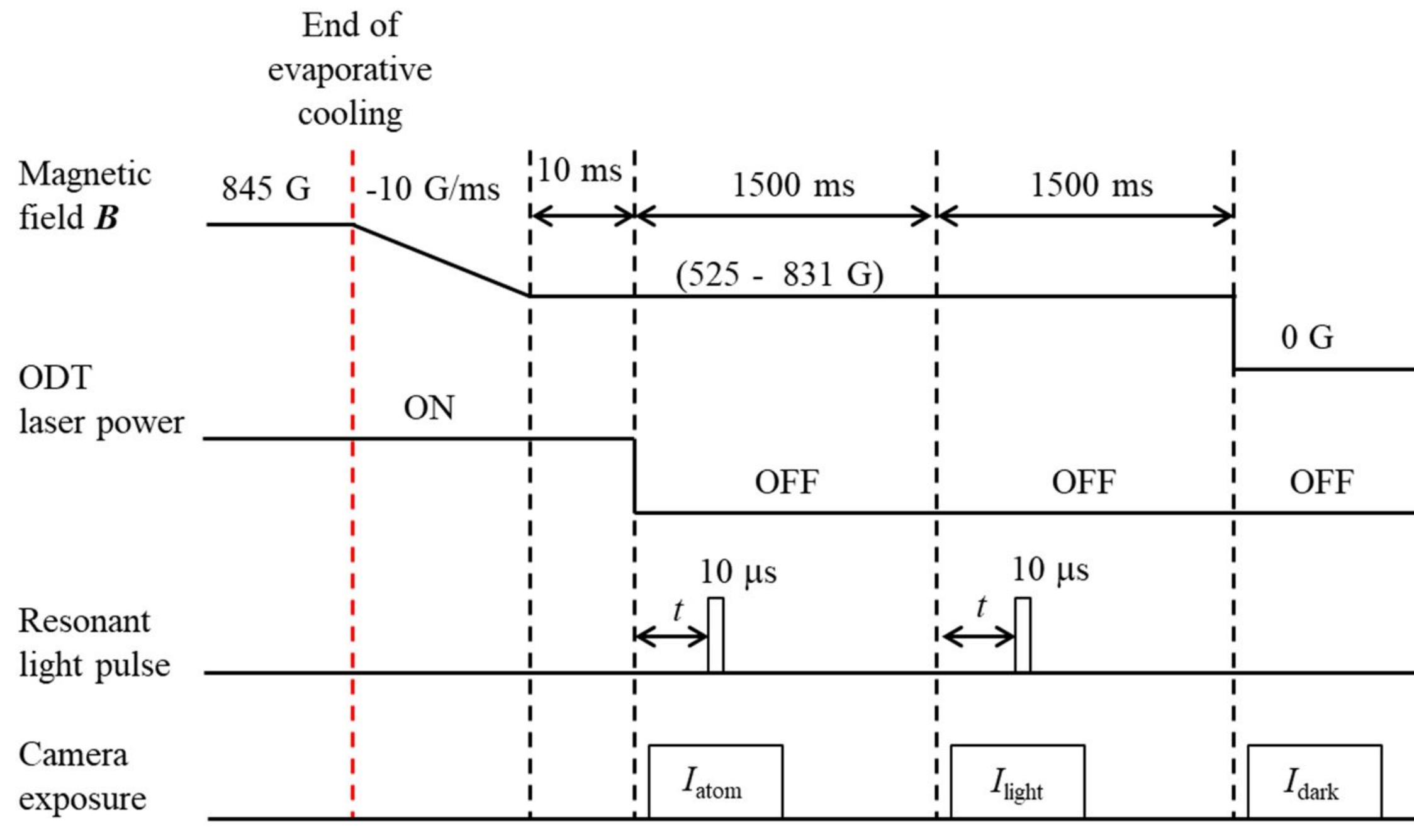}
	\caption{\textbf{Time sequence of the resonance absorption imaging (RAI)}\newline
    Absorption imaging is conducted at non-zero magnetic field $B$ ranging from $527$~G to $841$~G. The ODT laser power is switched off by Acoustic Optical Modulator (AOM) in less than $1~\mu$s.}
	\label{fig:figS3}
\end{figure}

Figure~\ref{fig:figS3} illustrates the time sequence of the resonance absorption imaging (RAI). At the end of forced evaporative cooling, the magnetic field is ramped down from $B=841$~G to the set value $B$ ($527$--$841$~G) at a rate of $10$~G/ms and stabilized for $10$~ms. The ODT laser power is then switched off and the Fermi gas is released to expand. At a given expansion time $t$, the gas is shined by a resonant light pulse of $10~\mu$s and the absorption image $I_{\rm atom}(x,z)$ is recorded by the EMCCD camera. Two more images are recorded, each at $1500$~ms later. Because the RAI technique is destructive, the second image $I_{\rm light}(x,z)$ is recorded as a reference with no absorption. The third image $I_{\rm dark}(x,z)$ is recorded with all light blocked to serve as the dark counts of the EMCCD camera.
The recorded images are digitized into two-dimensional (2D) matrices to obtain the optical density profile ${\rm OD}(x,z)$, which is computed with,
\begin{equation}
{\rm OD}(x,z)=\ln\left(\frac{I_{\rm light}-I_{\rm dark}}{I_{\rm atom}-I_{\rm dark}}\right)
\label{eq:(C1)}\,.
\end{equation}
The 2D atom number density profile for the recorded spin state $|1\rangle$ is then derived using the equation 
\begin{equation}
    \hat{\rho}(x,z)={\rm OD}(x,z) A_{\rm eff}/\sigma_{0}\,,
\end{equation}
where $A_{\rm eff}=9.81~\mu{\rm m}^{2}$ is the effective area of each pixel after magnification, and $\sigma_{0}=0.143~\mu{\rm m}^{2}$ is the resonant absorption cross section. When the optical density drops below unity, signal-to-ratio of the absorption image is too poor to yield reliable measurement.

\subsection*{Gaussian fit}\label{si:fit}
\begin{figure}[hbt]
    \centering
	\includegraphics[width=0.95\linewidth]{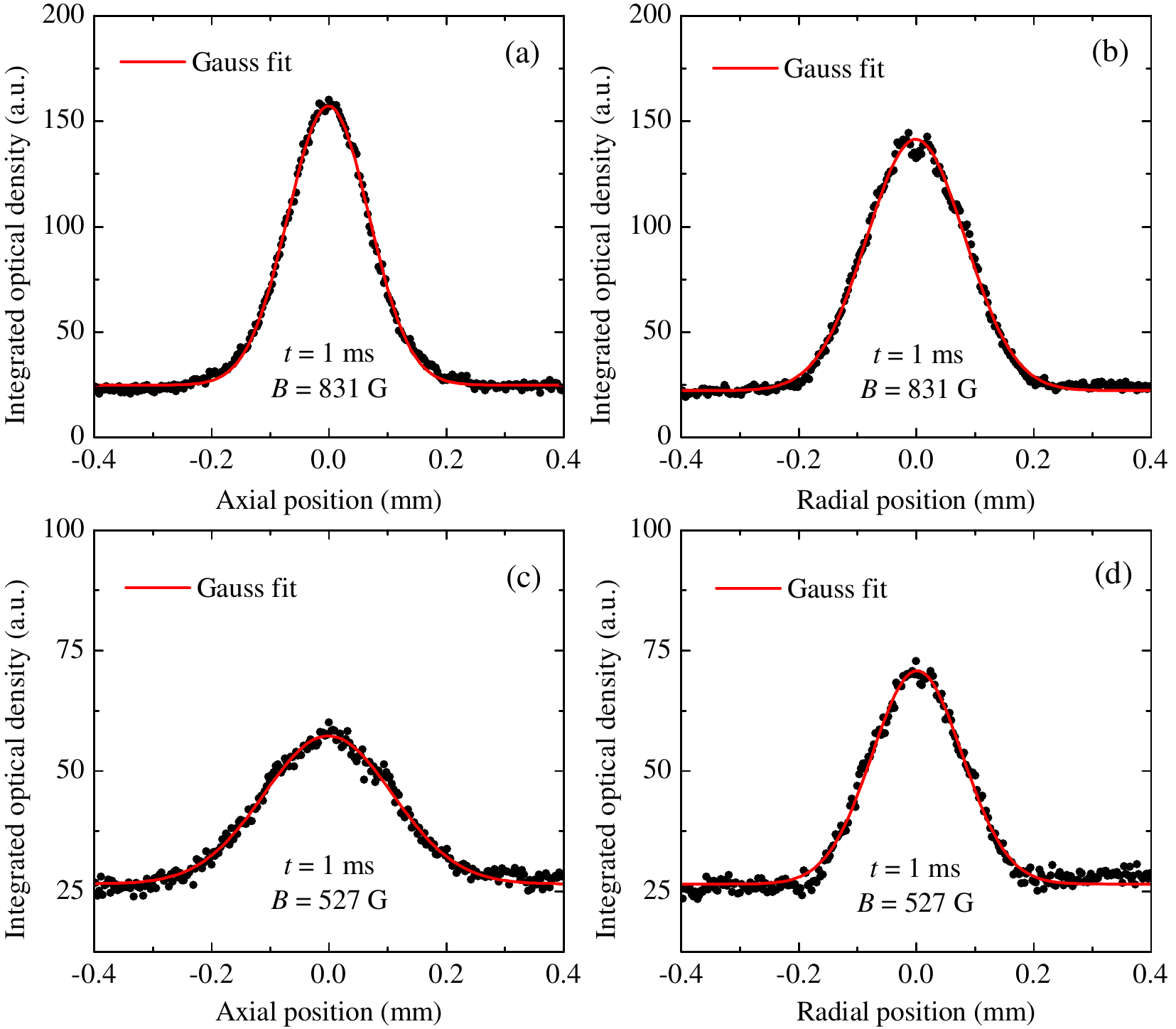}
	\caption{\textbf{Optical density (OD) profiles and Gaussian fits}\newline
    Axial and radial projections of the OD profiles of cold Fermi gases at time-of-flight (TOF) of 1 ms after their release from potential traps are shown for the strongly interacting case at $B=831$~G in (a) and (b), and for the non-interacting case at $B=527$~G in (c) and (d). Curves in red represent Gaussian fits, and the relative statistical uncertainties on the fitted width are $0.4\%$, $0.6\%$, $1.0\%$ and $0.9\%$ for (a), (b), (c) and (d), respectively.} 
	\label{fig:figS4}
\end{figure}

 In our experiment ($T/T_{F}>0.5$), the density profile of the expanding gas can be well described with a three dimensional Gaussian function, 
 \begin{equation}
 \rho\left(x,y,z \right)= \rho_{0}\exp \left[-\left(\frac{x^{2}}{2\sigma_{x}^2}+\frac{y^{2}}{2\sigma_{y}^2}+\frac{z^{2}}{2\sigma_{z}^2} \right)  \right]\,,
 \label{eq.(E1)}
 \end{equation}
where $\rho_{0}$=
$\frac{N}{(2\pi)^{3/2}\sigma_{x}\sigma_{y}\sigma_{z}}$ is the center density of the cold cloud~\cite{PhysRevA.55.4346}. 
The RAI process produces a 2D matrix of optical density ${\rm OD}(x,y)$, and it represents the column density profile within the imaging plane perpendicular to the line-of-sight. The matrix is further integrated along axial or radial direction to produce one-dimensional (1D) density profiles. As shown in Fig.~\ref{fig:figS4}, Gaussian fit was applied to the 1D density profiles and yields the mean root squared (rms) radius ($\sigma_{x}$ and $\sigma_{z}$) of the expanding cloud and a high coefficient of determination (R-square)  $\sim 0.99$.

\subsection*{Ballistic expansion and temperature measurement}\label{si:ballistic}
According to~\cite{Menotti2002}, the scaled RMS size $b_{i}=\sigma_{i}(t)/\sigma_{i}(0)$ of ballistically expanding gas fulfills a simple differential equation, $b_{i}^{''}(t)-\omega_{i}^{2}/b_{i}(t)^{3}=0$, where $\omega_{i}$ is the trapping frequency along $i$-axis. This equation has an analytical solution $b_{i}(t)=\sqrt{1+(\omega_{i}t)^{2}}$ at the given initial condition, $b_{i}(0)=1$ and $b_{i}^{'}(0)=0$. The radius $\sigma_{i}(t)$ is thus described by $\sigma_{i}(t)=\sigma_{i}(0)\sqrt{1+(\omega_{i}t)^{2}}$, where $\sigma_{i}(0)=(k_{B}T/m\omega_{i}^{2})^{1/2}$ is the initial in-trap RMS radius. 
This can be rewritten into Eq.~(1) 
in the main text, where $\left\langle v_{i}^{2}\right\rangle =k_{B}T_{i}/m$. 
Equation~(1) in the main text 
is widely used as a fitting function to extract temperature of non-interaction cold atom systems~\cite{JOSA1989Molasses,Menotti2002}.

In our experiment, the confining potential is ellipsoidal with $\omega_{x} \sim \omega_{y}=\omega_{r}$ and $\beta = \omega_{z} / \omega_{r} < 1$; $\beta$ determines the in-trap aspect ratio. The aspect ratio of the ballistically expanding Fermi gas is then given by~\cite{Menotti2002}
\begin{equation}
	\frac{\sigma_{r}(t)}{\sigma_{z}(t)}= \frac{\omega_{z}}{\omega_{r}} \frac{\sqrt{1+(\omega_{r}t)^{2}}}{\sqrt{1+(\omega_{z}t)^{2}}}
	\label{eq:(7)} \,.
\end{equation}
According to Eq.~\ref{eq:(7)}, the aspect ratio of the expanding non-interacting Fermi gas approaches unity asymptotically.

As shown in Fig.~\ref{fig:figS1}(b), the $s$-wave scattering length $a_{s}$ vanishes at $B=527$~G. At the end of evaporative cooling, the magnetic field is tuned to $527$~G, and then the ODT laser light is switched off abruptly. The released gas expands in the homogeneous magnetic field stabilized at $527$~G. Since the gas is non-interacting and follows ballistic expansion, the temperature can be extracted by fitting Eq.~(1) 
in the main text to the size of the expanding gas from a series of TOF images, as shown in Fig.~2(b) 
in the main text. The temperature parameters obtained from expansion data in the $x$ and $z$ directions are similar, $T_{x} = 4.56~\mu$K and $T_{z} = 4.71~\mu$K; the momentum distribution of the released Fermi gas is indeed isotropic. The mean value $T=(2T_{x}+T_{z})/3$ is used as the temperature $T$ in Table~\ref{tab:gas}. 

It should be noted that the expansion time range that can be used for temperature fitting is limited. For short expansion time, center atom density of the expanding gas is still high such that the absorption image is strongly saturated, resulting in an artificially larger RMS size extracted from Gaussian fitting. Moreover, the gas size is small at short expansion time, comparable to the spatial resolution of the imaging system. As a result the diffraction effect is significant in the radial direction ($x$-axis shown in Fig.~\ref{fig:figS2}), such that the fitted RMS size is also artificially larger than truth. Last, a strong loss of atoms is unavoidable when $B$ is tuned across $650$~G; it is found that the atom number drops by half after $B$ is swept to $527$~G. This puts a limitation on the longest applicable expansion time for imaging with high enough signal-to- ratios. Based on the above considerations, only data points within the range $0.3$--$1.3$~ms are chosen for temperature fitting at $B=527$~G.

\subsection*{Anisotropic expansion}\label{si:expansion}
\begin{figure}[hbt]
    \includegraphics[width=\linewidth]{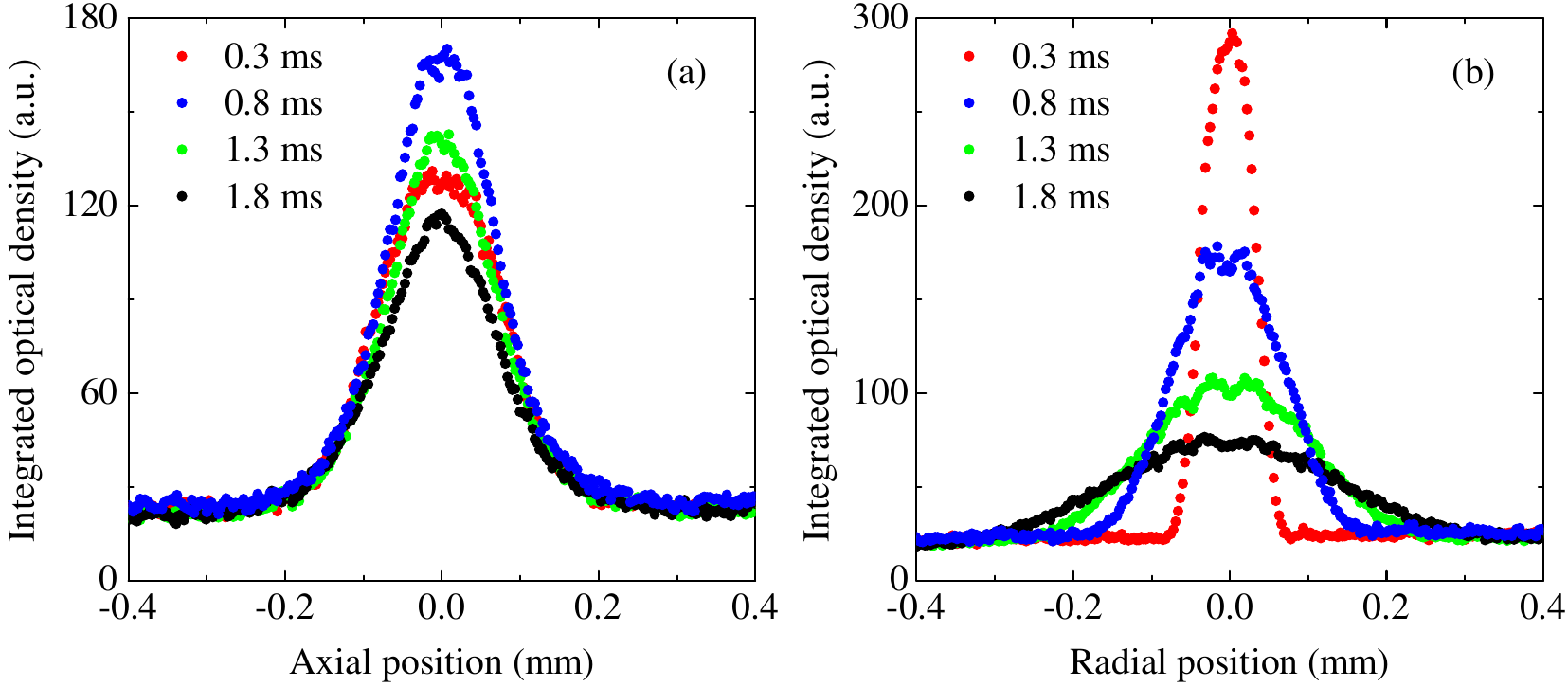}
	\caption{\textbf{Optical density (OD) profiles at different expansion times}\newline 
    Integrated one-dimensional OD profiles are shown at different expansion times, $0.3$~ms (red), $0.8$~ms (blue), $1.3$~ms (green) and $1.8$~ms (black), along axial (a) and radial direction (b). Because of absorption saturation, the axial profile of the center density at 0.3 ms is lower than that at $0.8$~ms and $1.3$~ms. These profiles are obtained at $B = 831$~G.}
	\label{fig:figS5}
\end{figure}

At the end of evaporative cooling, the ODT laser power is reduced to $1$~W per beam, and the gas is in normal state with a temperature above one half the Fermi temperature ($T>0.5T_{F}$).  When $B$ is tuned to the vicinity of the Feshbach resonance point ($B = 834$~G), the interaction strength parameter $|k_{F} a_{s}| \gg 1$, and an anisotropic expansion can be observed once the gas is released. This can be treated as hydrodynamic expansion where the positive $s$-wave scattering length is extremely large and the collisional cross section $\sigma_{s}$ (see Eq.~\ref{eq:(B1)}) is unitary limited~\cite{Menotti2002, Giorgini2008}. Whereas the anisotropic expansion in Ref.(17) is observed at large and negative $s$-wave scattering length ($a_{s}\sim-10^{4}a_{0}$) and the gas is highly degenerate ($T\sim 0.1T_{F}$). Here, $a_{0}=0.53\times 10^{-10}$~m is the Bohr radius. Figure~\ref{fig:figS5} shows one-dimensional optical density profile at different expansion times. The external magnetic field is set to $B=831$~G, where the $s$-wave scattering length is positive ($a_{s}=18.6~\mu$m) and extremely large (compared to the mean inter-atomic space of the cold cloud). The gas expands rapidly in the radial direction (Fig.~\ref{fig:figS5}(b)) while remains nearly stationary in the axial direction (Fig.~\ref{fig:figS5}(a)) in the measured time range from $0.3$~ms to $1.8$~ms. The RMS sizes of the expanding gas are extracted from Gaussian fits to the density profile, as shown in Fig.~2 
in the main text. 
Similar exercises are performed with other $B$ values. We find that the coupled nonlinear equations for superfluid expansion are not applicable to the observed anisotropic expansion of normal state Fermi gas~\cite{Ohara2002Science,Giorgini2008}.
We defer the study of anisotropic expansion of cold atom gases in the superfluid phase to a future work.

\begin{figure}
    \centering
    \includegraphics[width=0.5\textwidth]{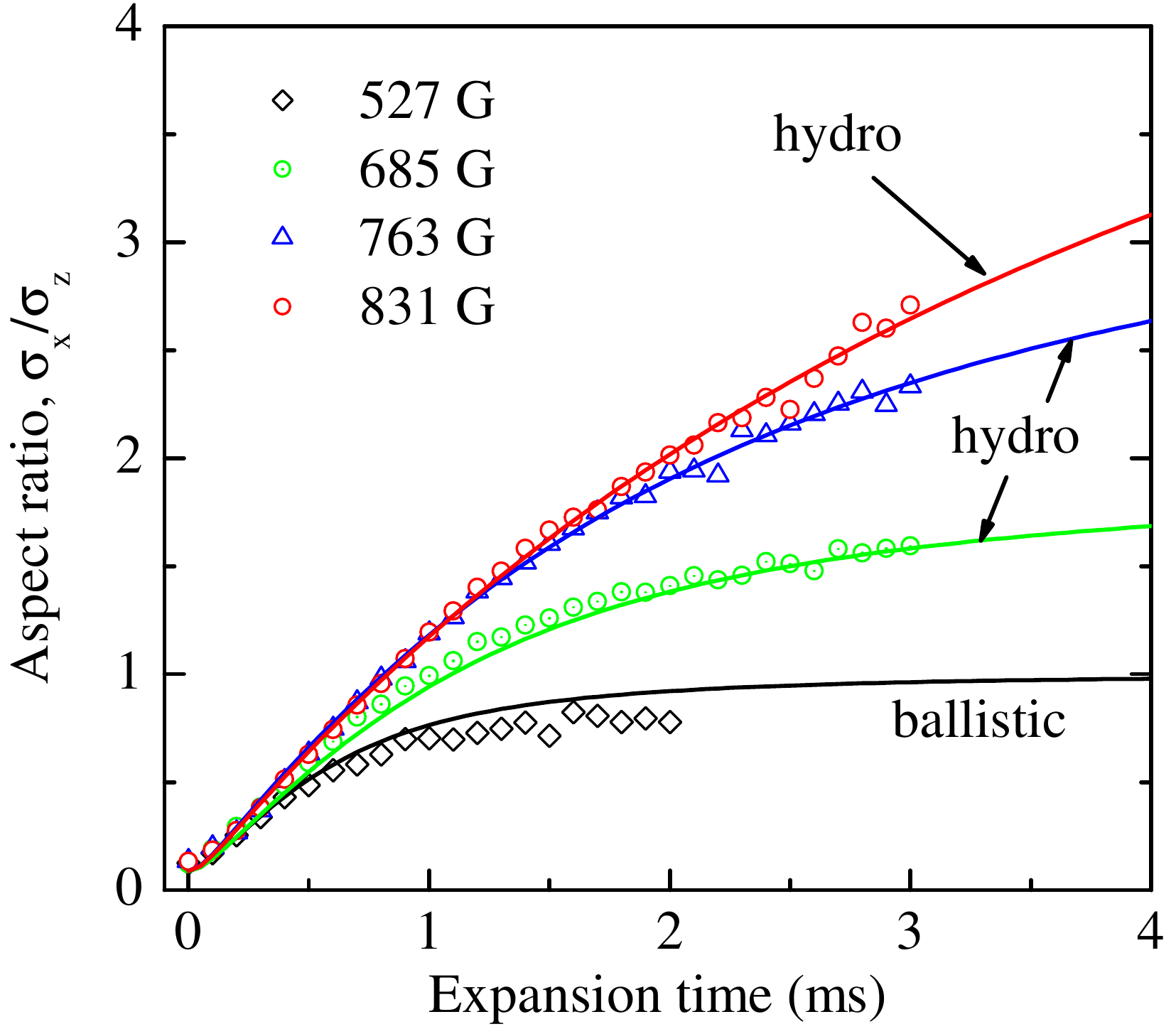}
    \caption{\textbf{Aspect ratio of the expanding cold atom $^{6}$Li Fermi gas}\newline
    The aspect ratio of the expanding cold atom $^{6}$Li Fermi gas is plotted as a function of expansion time. The points are experimental data. The solid curves are theoretical calculations of ballistic expansion (black) and collisional hydrodynamic fits (blue, green, red) with different $\chi$ values. See text for details.}
    \label{fig:figS6}
\end{figure}

Figure~\ref{fig:figS6} shows the aspect ratio as a function of expansion time for four $B$ values. At $B=527$~G, the expansion is ballistic and the aspect ratio does not reverse and never exceeds unity. When the magnetic field is set at $685$~G, a moderate reversion of the aspect ratio is observed. When the magnetic field is tuned to values around the Feshbach resonance, namely $763$ and $831$~G, the aspect ratio quickly exceeds unity at expansion time shorter than $1$~ms. The solid curves in Fig.~\ref{fig:figS6} are theoretical calculations with a set of coupled nonlinear equations as described in Ref (23) in the main text,
\begin{equation}
	b_{i}^{''}(t)-\frac{\omega_{i}^{2}}{b_{i}^{3}(t)}+\frac{3}{2}\chi \omega_{i}^{2}\left( \frac{1}{b_{i}^{3}(t)}-\frac{1}{b_{i}(t)V(t)}\right) =0
	\label{eq:(S3)}\,,
\end{equation}
where $\chi$ quantifies the interaction strength, and $V\left( t\right) =b_{x}\left( t\right) b_{y}\left( t\right) b_{z}\left( t\right)$ is the scaled volume. For ballistic expansion at $B=527$~G, $\chi=0$; while for $B=685$, $763$ and $831$~G, $\chi=0.5$, $0.63$ and $0.66$ is chosen respectively. The high $\chi$ value represents a strong interaction strength within the gas when $B$ is tuned to the vicinity of Feshbach resonance point ($B=834$~G). These comparisons support the collisional hydrodynamic picture of anisotropic expansion of Fermi gas at large interaction strengths in normal state.

\subsection*{Treatment of statistical uncertainties}\label{si:error}
The trapped atomic ensemble rapidly expands upon release, with its central density decreasing 100-fold within $2$~ms. This rapid expansion causes absorption images to blur with relatively growing background, particularly at longer expansion times. Consequently, measured cloud sizes and atom numbers become increasingly scattered and systematically underestimated, as evidenced by data points deviating from fitting curves in Fig.~2 
in the main text. We thus restrict our analysis to a $1$~ms time window where the atom number decreases by $<30$\% and peak OD remains between $0.5$--$2$.
For the experiments with cross angle of $10^\circ$, the fit range is kept at $0.3$--$1.3$~ms for week interaction ($B<630$~G) and $1.0$--$2.0$~ms for stronger interaction ($B>630$~G). For those with cross angle of $35^\circ$, 
the fitting range for all interaction is fixed at $0.5$--$1.5$~ms. 
The uncertainties $\delta\langle{v}_{x}^{2}\rangle$ and $\delta\langle{v}_{z}^{2}\rangle$ on the fitted velocities are the standard fit uncertainties, which are determined by those on $\sigma_x$ and $\sigma_z$ data points shown in Fig.~2 
in the main text. 

We have chosen the fixed duration of $1$~ms for all fit ranges in obtaining the velocities to keep uniformity.
Varying the size of this common duration, and sliding the fit ranges to earlier or later times keeping the fit range size fixed, do not significantly affect the $v_2$ values.

The uncertainties on $v_2$ are obtained from standard error propagation by
\begin{equation}
   \delta{v}_{2}= \frac{1-v_{2}^{2}}{2}\sqrt{
   \left(\frac{\delta\langle{v}_{x}^{2}\rangle}{\langle{v}_{x}^{2}\rangle}\right)^{2} + \left(\frac{\delta\langle{v}_{z}^{2}\rangle}{\langle{v}_{z}^{2}\rangle}\right)^{2} }
   \label{eq:(B3)}\,.
\end{equation}
These uncertainties are plotted in Figs.~3, 4(a), and 5 
in the main text.


\subsection*{Estimate of shear viscosity}
The ratio of shear viscosity over entropy density, $\eta/s$, is an important property describing fluids. 
It is conjectured to be limited by a lower bound of $1/4\pi$ (in the unit of $\hbar/k_B$) from string theory~\cite{Kovtun:2004de}. 
The $\eta⁄s$ of the QGP was found to be close to this quantum limit, leading to its conclusion of being a  nearly perfect fluid~\cite{Gyulassy:2004zy,ThomasPT2010}.

According to the Cao {\em et al.}~\cite{Cao2011} the shear viscosity can be estimated by 
$\eta\approx 2.77\rho\left(T/T_{F}\right)^{3/2}$ for a atom gas of number density $\rho$ at temperature relevant for our experiment, $T/T_{F}\approx 0.7$.
For unitary Fermi gas, such a form arises from hydrodynamics to determine the curvature of the aspect ratio versus time. 
The entropy density of a unitary Fermi gas can be estimated by 
$\sim 2.7\rho$~\cite{PhysRevLett.98.080402}. 
The shear viscosity to entropy density ratio is therefore $\eta/s\approx\ \left(T/T_{F}\right)^{3/2}\approx0.6$ for our Fermi gas experiment.
This is in line with that in Ref.~\cite{Cao2011} and is a few times the lower quantum limit~\cite{Kovtun:2004de}.
The shear viscosity is typically given by the ratio of the momentum scale to the interaction cross section $\eta \sim p/\sigma$. 
For our atom gas with a lower interaction strength (lower opacity), $\eta/s$ increases, inversely proportional to the opacity. 





\end{document}